%% file: main_and_supp.tex
\documentclass[superscriptaddress,amsmath,amssymb,aps,prb,twocolumn,nofootinbib]{revtex4-2}
\usepackage{graphicx}         
\usepackage{dcolumn}          
\usepackage{bm}               
\usepackage{upgreek}
\usepackage{booktabs} 				
\usepackage{tabularx}
\usepackage{array}
\usepackage[colorlinks=true,urlcolor=blue,breaklinks=true]{hyperref}

\usepackage{xcolor}

\newcommand{\sign}{\mathrm{sign}}

\newcommand{\total}{\mathrm{tot}}
\newcommand{\lef}{\left(}
\newcommand{\rig}{\right)}

\newcommand{\Le}{\textsc{L}}
\newcommand{\Ri}{\textsc{R}}

\newcommand{\kB}{k_\textsc{b}}

\newcommand{\Fermi}{\textsc{f}}
\newcommand{\Dirac}{\textsc{d}}
\newcommand{\Bohr}{\textsc{b}}
\newcommand{\side}{\mathrm{leg}}
\newcommand{\cycl}{\mathrm{g}}
\newcommand{\cont}{\textsc{c}}

\newcommand{\SR}{\mathcal{SR}}
\newcommand{\NW}{\textsc{nw}}

\newcommand{\vecB}{\mathbf{B}}
\newcommand{\vecF}{\mathbf{F}}
\newcommand{\vecv}{\mathbf{v}}
\newcommand{\vecu}{\mathbf{u}}
\newcommand{\veck}{\mathbf{k}}
\newcommand{\vece}{\mathbf{e}}

\newcommand*{\nolink}[1]{%
    #1
}

\begin{document}

\title{In-plane magnetic field-driven symmetry breaking \\ in topological insulator-based three-terminal junctions}

\author{Jonas K\"{o}lzer}
\email{j.koelzer@fz-juelich.de}
\affiliation{Peter Gr\"unberg Institut (PGI-9), Forschungszentrum J\"ulich, 52425 J\"ulich, Germany}
\affiliation{JARA-Fundamentals of Future Information Technology, J\"ulich-Aachen Research Alliance, Forschungszentrum J\"ulich and RWTH Aachen University, Germany}

\author{Kristof Moors}
\email{k.moors@fz-juelich.de}
\affiliation{Peter Gr\"unberg Institut (PGI-9), Forschungszentrum J\"ulich, 52425 J\"ulich, Germany}
\affiliation{JARA-Fundamentals of Future Information Technology, J\"ulich-Aachen Research Alliance, Forschungszentrum J\"ulich and RWTH Aachen University, Germany}

\author{Abdur Rehman Jalil}
\affiliation{Peter Gr\"unberg Institut (PGI-9), Forschungszentrum J\"ulich, 52425 J\"ulich, Germany}
\affiliation{JARA-Fundamentals of Future Information Technology, J\"ulich-Aachen Research Alliance, Forschungszentrum J\"ulich and RWTH Aachen University, Germany}

\author{Erik Zimmermann}
\affiliation{Peter Gr\"unberg Institut (PGI-9), Forschungszentrum J\"ulich, 52425 J\"ulich, Germany}
\affiliation{JARA-Fundamentals of Future Information Technology, J\"ulich-Aachen Research Alliance, Forschungszentrum J\"ulich and RWTH Aachen University, Germany}

\author{Daniel Rosenbach}
\affiliation{Peter Gr\"unberg Institut (PGI-9), Forschungszentrum J\"ulich, 52425 J\"ulich, Germany}
\affiliation{JARA-Fundamentals of Future Information Technology, J\"ulich-Aachen Research Alliance, Forschungszentrum J\"ulich and RWTH Aachen University, Germany}

\author{Lidia Kibkalo}
\affiliation{Ernst Ruska-Centre for Microscopy and Spectroscopy with Electrons, Materials Science and Technology, Forschungszentrum J\"ulich, 52425 J\"ulich, Germany}

\author{Peter Sch\"{u}ffelgen}
\affiliation{Peter Gr\"unberg Institut (PGI-9), Forschungszentrum J\"ulich, 52425 J\"ulich, Germany}
\affiliation{JARA-Fundamentals of Future Information Technology, J\"ulich-Aachen Research Alliance, Forschungszentrum J\"ulich and RWTH Aachen University, Germany}

\author{Gregor Mussler}
\affiliation{Peter Gr\"unberg Institut (PGI-9), Forschungszentrum J\"ulich, 52425 J\"ulich, Germany}
\affiliation{JARA-Fundamentals of Future Information Technology, J\"ulich-Aachen Research Alliance, Forschungszentrum J\"ulich and RWTH Aachen University, Germany}

\author{Detlev Gr\"utzmacher}
\affiliation{Peter Gr\"unberg Institut (PGI-9), Forschungszentrum J\"ulich, 52425 J\"ulich, Germany}
\affiliation{JARA-Fundamentals of Future Information Technology, J\"ulich-Aachen Research Alliance, Forschungszentrum J\"ulich and RWTH Aachen University, Germany}

\author{Thomas L. Schmidt}
\affiliation{Department of Physics and Materials Science, University of Luxembourg, L-1511 Luxembourg}

\author{Hans L\"uth}
\affiliation{Peter Gr\"unberg Institut (PGI-9), Forschungszentrum J\"ulich, 52425 J\"ulich, Germany}
\affiliation{JARA-Fundamentals of Future Information Technology, J\"ulich-Aachen Research Alliance, Forschungszentrum J\"ulich and RWTH Aachen University, Germany}

\author{Thomas Sch\"apers}
\email{th.schaepers@fz-juelich.de}
\affiliation{Peter Gr\"unberg Institut (PGI-9), Forschungszentrum J\"ulich, 52425 J\"ulich, Germany}
\affiliation{JARA-Fundamentals of Future Information Technology, J\"ulich-Aachen Research Alliance, Forschungszentrum J\"ulich and RWTH Aachen University, Germany}

\hyphenation{}
\date{\today}

\input{abstract.tex}

\maketitle

\input{main_text.tex}

\section*{Acknowledgments}
This work was partly funded by the Deutsche Forschungsgemeinschaft (DFG, German Research Foundation) under Germany’s Excellence Strategy – Cluster of Excellence Matter and Light for Quantum Computing (ML4Q) EXC 2004/1 – 390534769, by the German Federal Ministry of Education and Research (BMBF) via the Quantum Futur project "MajoranaChips" (Grant No.\ 13N15264) within the funding program Photonic Research Germany, and by the Bavarian Ministry of Economic Affairs, Regional Development and Energy within Bavaria’s High-Tech Agenda Project “Bausteine für das Quantencomputing auf Basis topologischer Materialien mit experimentellen und theoretischen Ansätzen” (grant allocation no.\ 07 02/686 58/1/21 1/22 2/23).

\section*{Author contributions}
J.K., A.R.J., E.Z.\ and D.R.\ were involved in the sample fabrication, A.R.J., P.S.\ and G.M.\ grew the thin film using MBE, J.K., E.Z.\ and D.R.\ performed the low temperature transport measurements, L.K.\ prepared the lamella and did the high-resolution scanning transmission electron microscopy analysis, K.M.\ performed the simulations and J.K.\ and K.M.\ analyzed the data. K.M., J.K., T.S., T.L.S.\ and H.L.\ wrote the paper with contributions from all co-authors, the project was supervised by D.G., T.L.S., H.L.\ and T.S.\ and all authors contributed to the discussions.

\input{main.bbl}
\clearpage
\widetext

\setcounter{section}{0}
\setcounter{equation}{0}
\setcounter{figure}{0}
\setcounter{table}{0}
\setcounter{page}{1}
\makeatletter
\renewcommand{\thesection}{S\Roman{section}}
\renewcommand{\thesubsection}{\Alph{subsection}}
\renewcommand{\theequation}{S\arabic{equation}}
\renewcommand{\thefigure}{S\arabic{figure}}
\renewcommand{\figurename}{Supplementary Figure}
\renewcommand{\bibnumfmt}[1]{[S#1]}
\renewcommand{\citenumfont}[1]{S#1}

\begin{center}
\textbf{Supplemental Material for \\ In-plane magnetic field-driven symmetry breaking \\ in topological insulator-based three-terminal junctions}
\end{center}

\input{supp_text.tex}

\input{supp.bbl}
\end{document}

%% file: abstract.tex
\begin{abstract}
\section*{Abstract}
Topological surface states of three-dimensional topological insulator nanoribbons and their distinct magnetoconductance properties are promising for topoelectronic applications and topological quantum computation. A crucial building block for nanoribbon-based circuits are three-terminal junctions. While the transport of topological surface states on a planar boundary is not directly affected by an in-plane magnetic field, the orbital effect cannot be neglected when the surface states are confined to the boundary of a nanoribbon geometry. Here, we report on the magnetotransport properties of such three-terminal junctions. We observe a dependence of the current on the in-plane magnetic field, with a distinct steering pattern of the surface state current towards a preferred output terminal for different magnetic field orientations. We demonstrate that this steering effect originates from the orbital effect, trapping the phase-coherent surface states in the different legs of the
junction on opposite sides of the nanoribbon and breaking the left-right symmetry of the transmission across the junction. The reported magnetotransport properties demonstrate that an in-plane magnetic field is not only relevant but also very useful for the characterization and manipulation of transport in three-dimensional topological insulator nanoribbon-based junctions and circuits, acting as a topoelectric current switch. 
\end{abstract}

%% file: main_text.tex
\renewcommand{\figurename}{Figure}

\section*{Introduction} \label{sec:introduction}
The behavior of spin-momentum-locked surface states in 3D~TI-based multiterminal junctions is crucial for their use in topoelectronic and spintronic circuit applications, and Majorana-based topological quantum computation architectures \cite{Moore2010,Alicea2011,VanHeck2012,Hyart13,Aasen2016,Litinski2017,Schueffelgen2019}. In past studies, various transport properties of straight 3D~TI-based nanowires and ribbons have been investigated theoretically and observed experimentally in micrometer- and nanometer-sized systems, e.g., weak antilocalization and quasiballistic transport with Aharonov--Bohm oscillations~\cite{Peng10,Bardarson10,Xiu11,Bardarson2013,Dufouleur13,Jauregui16,Arango16,Dufouleur17,Ziegler18,Xypakis2020,Rosenbach20}.

In addition to magnetotransport studies, first steps have been made to use this platform for hosting exotic quasiparticle states known as Majorana bound states (MBSs), by aligning a 3D~TI nanowire with an external magnetic field and combining it with an $\mathit{s}$-wave superconductor for realizing topological superconductivity via the proximity effect~\cite{Cook11, Cook2012, Schueffelgen19b, Liu19, Bai2020}. These states are of particular interest since they are promising candidates for the realization of fault-tolerant quantum computation~\cite{Ivanov01, Kitaev_2003, Freedman_2003, Hassler2011, Hyart13, Litinski2017}. By exploiting their nonlocal nature and nonabelian exchange statistics, it has been proposed that qubits and quantum operations can be implemented with MBSs in a very robust manner. A key operation for this approach is the braiding of different pairs of MBSs. In order to perform braiding, however, straight nanowire structures are not sufficient. More complex structures such as three-terminal junctions (referred to as tri-junctions below) and eventually networks of 3D~TI nanowires are required, and the magnetic field has to be aligned appropriately. In this regard, a proper understanding of the impact of an in-plane magnetic field on the electron transport of coherent topological surface states across such junctions is essential while, unlike on straight nanowires, experimental transport studies are still lacking.

It has already been predicted theoretically that the conductance in 3D~TI nanowire-based structures, such as kinks and Y-junctions, can be controlled by applying an in-plane magnetic field~\cite{Moors18}. The underlying reason for these conductance properties is that the relative orientation of the magnetic field and the junction affects the transmission to the different arms through the orbital effect. In the single-channel limit, quantum transport simulations indicate that a complete pinch-off or near-perfect transparency of the topological surface state-based carrier transport can be realized to particular output legs for certain magnetic field orientations and strengths.

In this context, we have studied the low-temperature magnetotransport properties of Bi$_2$Te$_3$-based tri-junctions with three nanoribbon legs. In these junctions, the current is injected in a single input leg and splits into the two remaining output legs. The structures were prepared by employing selective-area molecular beam epitaxy (MBE) and the conductance was measured as a function of the angle between the magnetic field and the input lead. We have found characteristic transmission patterns with alternating optimal transmission into one of the two output legs depending on the orientation of a magnetic field aligned parallel to the plane of the junction. To explain the observed features, we developed a qualitative tri-junction transmission model, based on our findings from semiclassical considerations and quantum transport simulations (with Kwant~\cite{Groth14}) of topological surface states in 3D~TI-based multiterminal junctions in the presence of an external magnetic field.

\begin{figure}
  \centering
  \includegraphics[width=0.40\textwidth]{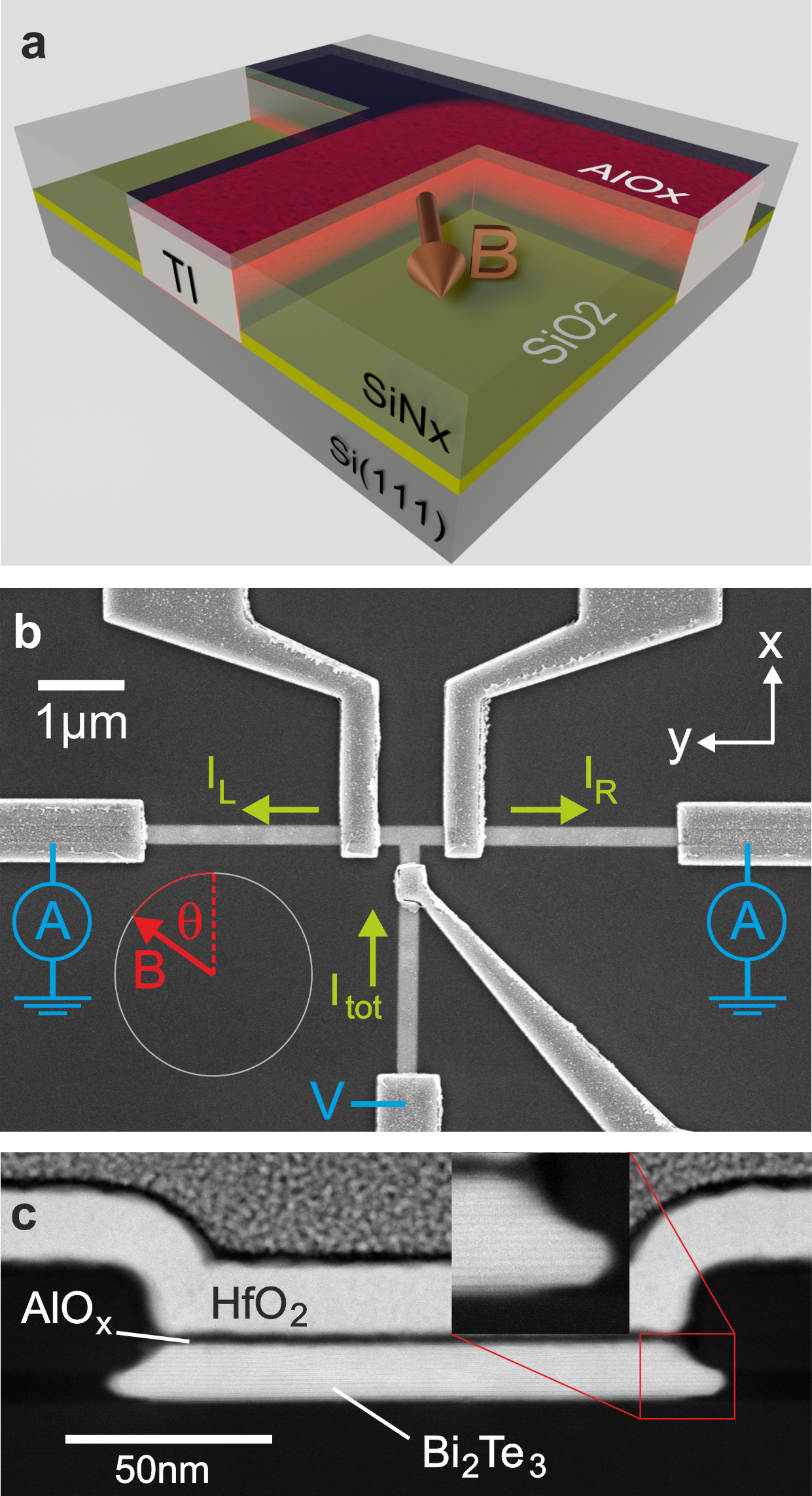}
  \caption{\textbf{Sample layout and microscopy images of the T-junction.}
    (\textbf{a}) Schematic illustration of the layout of a T-shaped selectively-grown 3D~TI nanostructure with in-plane magnetic field-driven steering effect indicated.
    (\textbf{b}) Scanning electron microscopy image of a T-junction device. The contacts near (far from) the T are referred to as the inner (outer) contacts. The magnetic field $\vecB$ is applied in-plane. The input current $I_\total$, which is injected from the outer contact on the bottom leg, splits off into currents $I_\Ri$ and $I_\Le$, which are measured at the outer contacts on the right and left leg of the T, respectively.
    (\textbf{c}) Transmission electron micrograph of the cross section of an AlO$_x$-capped Bi$_2$Te$_3$ nanoribbon. The scale bar indicates $50\,\textnormal{nm}$. The thickness of the ribbon is determined to be $14\,\textnormal{nm}$ and the average width is about $135\,\textnormal{nm}$. The stripes in the 3D~TI layer (see inset) correspond to quintuple layers of the material.
  }
  \label{fig:T-junction-SEM}
\end{figure}

\section*{Results} \label{sec:results}

\subsection*{Magnetoconductance properties} \label{subsec:total_conductance}
\begin{figure*}
  \centering
  \includegraphics[width=\linewidth]{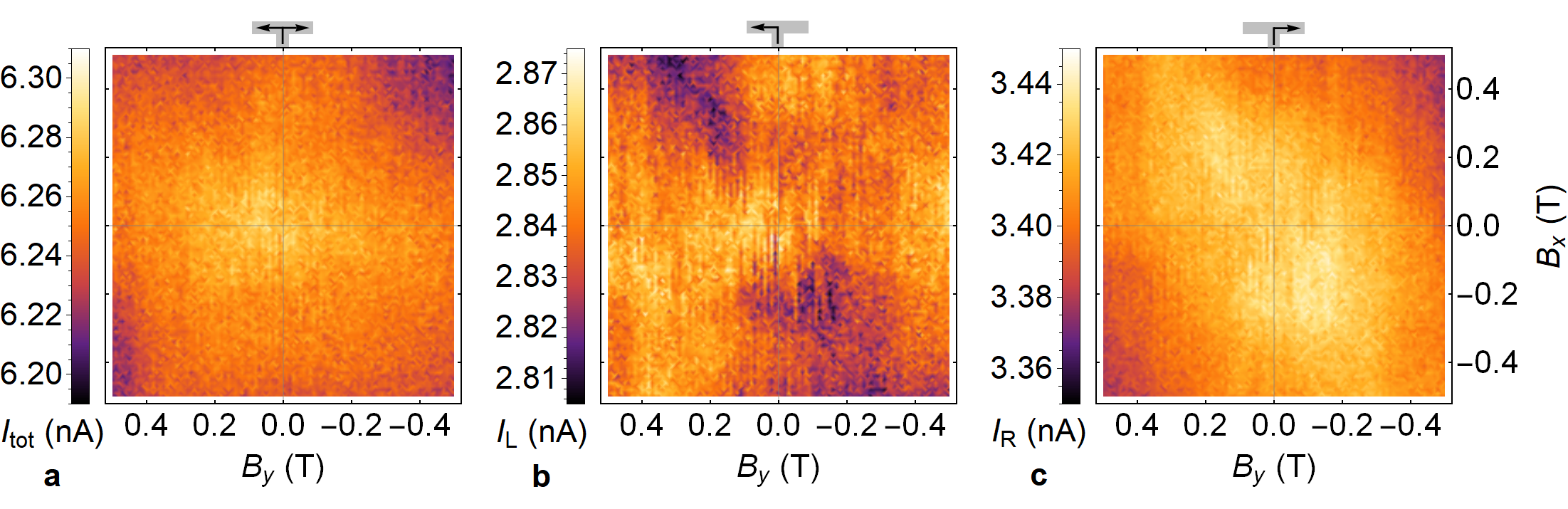}
  \caption{\textbf{Currents through the T-shaped tri-junction as a function of in-plane magnetic field components.}
      (\textbf{a})--(\textbf{c}) The current is measured from the bottom leg to both output legs (shown in \textbf{a}), and individually to the left leg (shown in \textbf{b}) and right leg (shown in \textbf{c}). The sketches on top of the plots indicate the current paths. The measurements were carried out at a temperature of 25\,mK for the T-shaped tri-junction presented in Fig.~\ref{fig:T-junction-SEM}b. The reference frame is also shown in that figure.
  }
  \label{fig:T-junction-current-map}
\end{figure*}

Low temperature magnetotransport measurements were performed on a T-shaped tri-junction. The sample is composed of a Bi$_2$Te$_3$ film at an average thickness of $14\,\textnormal{nm}$ that is grown selectively and capped with AlO$_x$ (see Fig.~\ref{fig:T-junction-SEM}a for layout and Methods section for details on growth and fabrication). The electron phase-coherence length below $T = 1\,\textnormal{K}$ was determined to be $l_{\phi} \approx 240\,\textnormal{nm}$ (see Methods section). Figs.~\ref{fig:T-junction-SEM}b and c show a scanning electron micrograph of the device and a scanning transmission electron micrograph of the cross section of one of the legs, respectively. The device is contacted by Ti/Au after selectively removing the capping and it is protected by an additional layer of HfO$_2$ grown by means of atomic layer deposition.
In our setup, we apply a voltage $V$ to the bottom terminal and ground the other two terminals while measuring the current flowing through each of the terminals as a function of the magnetic field applied in-plane.
Sweeping the in-plane field strength up to 0.5~T we observe a uniform decrease in the total current $I_\total$, which we attribute to the weak antilocalization (WAL) effect present in the individual legs of the T-junction (see Fig.~\ref{fig:T-junction-current-map}a). The total current is calculated as the sum of the currents going into the left and right leg of the T-junction. No pronounced orientation in the WAL pattern is observed. This result is in agreement with what has been measured on similar samples of straight nanoribbons in prior experiments~\cite{Rosenbach20, Weyrich19, Koelzer20}.

\subsection*{Steering ratio} \label{subsec:conductance}
\begin{figure}
  \centering
  \includegraphics[width=\linewidth]{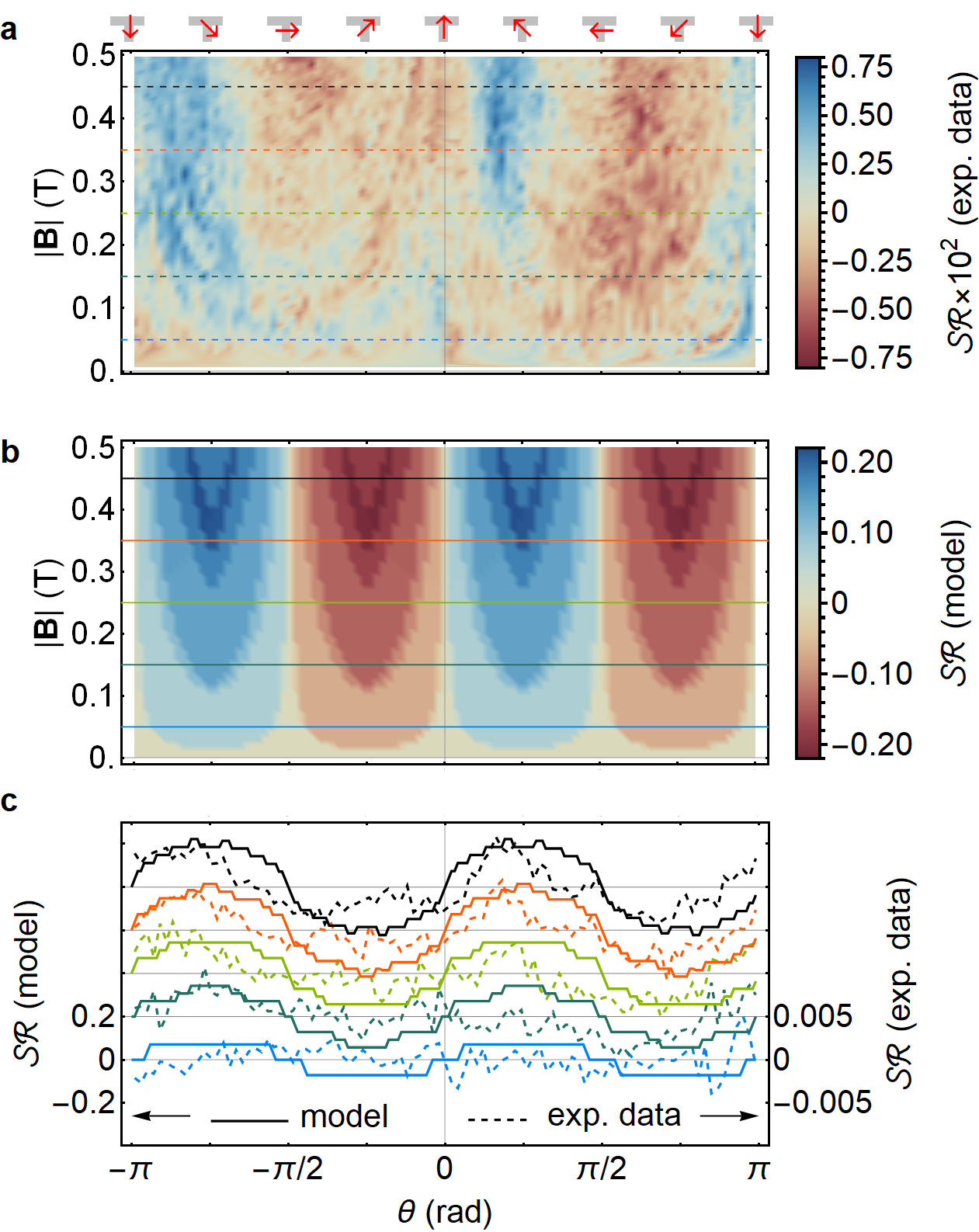}
  \caption{\textbf{Steering ratio of the T-shaped tri-junction.}
    (\textbf{a}),(\textbf{b}) The steering ratio $\SR$ of the T-shaped tri-junction presented in Fig.~\ref{fig:T-junction-SEM}b as a function of the in-plane magnetic field orientation $\theta$ (with the direction relative to the T-junction indicated by the red arrows at the top) and the field strength $|\vecB|$, as obtained from the transport measurements (see Fig.~\ref{fig:T-junction-current-map}) in \textbf{a}, and from the qualitative transmission model, as discussed in the Main Text and derived in Supplementary Note~\nolink{\ref{sec:details_model}}, in \textbf{b}.
    (\textbf{c}) Line cuts of the steering ratio for increasing magnetic field strengths, obtained from the experimental data (dashed lines) and model results (solid lines) presented in \textbf{a} and \textbf{b}, respectively. The scale indicated on the right (left) is for the experimental (model) values and the curves are shifted up by multiples of $0.5 \times 10^{-2}$ ($0.2$).
  }
  \label{fig:T-junction-B-constant}
\end{figure}

The individual currents are measured as a function of the in-plane magnetic field components and shown in Figs.~\ref{fig:T-junction-current-map}b--c. Unlike for the total current, a pronounced correlation along the diagonal ($B_x = B_y$) and antidiagonal ($B_x = - B_y$) of the $(B_x, B_y)$-plane can be identified for the current towards the right and the left, respectively. A robust steering pattern emerges over a large range of magnetic field strengths, with the current favoring one of the two output legs depending on the in-plane magnetic field orientation. The steering pattern is more clearly seen when plotting the steering ratio $\SR$, defined as
\begin{equation} \label{eq:steering_ratio}
    \SR = \frac{I_\Ri - I_\Le}{I_\Ri + I_\Le} - \left< \frac{I_\Ri - I_\Le}{I_\Ri + I_\Le} \right> _{|\mathbf{B}| = \textnormal{const.}},
\end{equation}
as a function of the magnetic field orientation angle and strength (see Fig.~\ref{fig:T-junction-B-constant}a, details on the transformation of the experimental data set can be found in the Methods section).
The pattern becomes more pronounced at higher field strengths and a $\pi$-periodicity as a function of the magnetic field orientation angle $\theta$, $\SR \propto \sin(2\theta)$, can easily be identified when taking a line cut for a fixed magnetic field strength (see Fig.~\ref{fig:T-junction-B-constant}c).

Note that the intrinsic, magnetic field-independent asymmetry of the junction is subtracted from the steering ratio in Equation~\eqref{eq:steering_ratio}. It is clear from the range of the individual currents in Fig.~\ref{fig:T-junction-current-map} that the measured currents already have some asymmetry in the absence of an external magnetic field. This asymmetry can be attributed to small structural differences in the wire legs and different contact resistances, for example, that are not directly related to the transmission of 3D~TI surface states across the tri-junction.
Further note that the angle dependence of the steering ratio immediately rules out an explanation based on the Hall effect due to a possible misalignment of the external magnetic field and sample planes, as it would give rise to a $2\pi$-periodic pattern. Other possible symmetry-breaking mechanisms that are unrelated to the tri-junction itself are considered in the Discussion section below and more details on them are provided in Supplementary Note~\nolink{\ref{sec:details_model}}.

\subsection*{Temperature dependence} \label{subsec:T-dependence}
\begin{figure}
  \centering
  \includegraphics[width=0.9\linewidth]{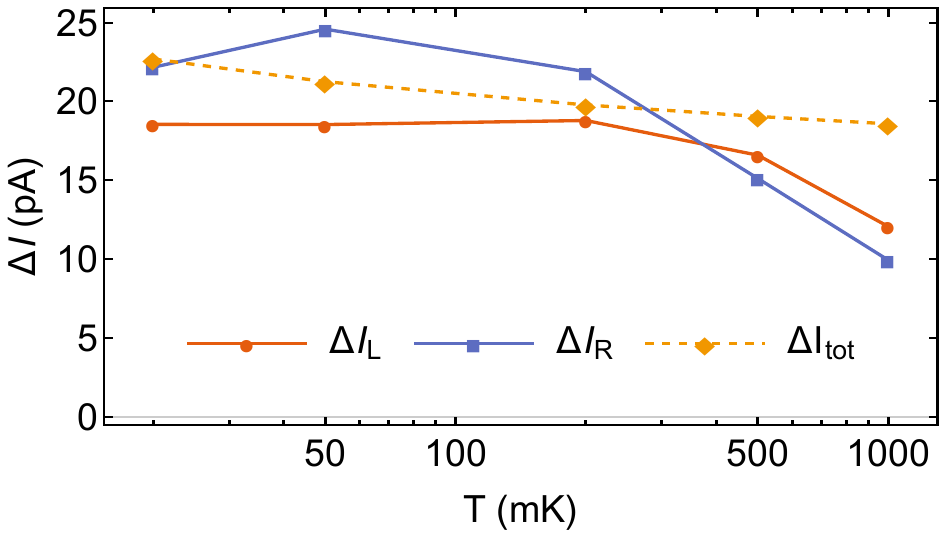}
  \caption{
    \textbf{The standard deviation of the current through a Y-shaped tri-junction}. The data is extracted from current data sets over different in-plane magnetic field strengths and orientations (see Supplementary Note~\nolink{\ref{sec:Y-junction}} for current data), evaluated at different temperatures.
  }
  \label{fig:Y-junction-T-dependence}
\end{figure}

Magnetotransport measurements have been conducted for a Y-shaped tri-junction at different temperatures to resolve the temperature dependence of the observed steering pattern (see Supplementary Note~\nolink{\ref{sec:Y-junction}} for the current measurement results). In general, we find that the dependence of the current on the magnetic field strength and orientation decreases for increasing temperature. This can be quantified with the standard deviation of the current data over all the measured magnetic fields, i.e., $-0.7\,\textnormal{T} \leq B_{x,y} \leq 0.7\,\textnormal{T}$, and comparing this quantity at different temperatures. The results obtained from the Y-junction magnetotransport data are shown in Fig.~\ref{fig:Y-junction-T-dependence}. The standard deviation of the total current displays a slow but steady decrease for increasing temperatures, which can be attributed to the change in the field strength dependence due to the effect of WAL near $|\vecB| = 0$. The standard deviation of the individual currents shows a very different temperature dependence. The standard deviation has a much steeper decrease above $T = 200\,\textnormal{mK}$ and appears to saturate below this temperature. This crossover in the profile coincides with the appearance of the steering pattern below $T \approx 200\,\textnormal{mK}$.

\subsection*{Origin of steering effect} \label{subsec:steering_effect}
The left-right symmetry breaking of the tri-junction and the steering effect of the current can be explained by considering the impact of the orbital effect on 3D~TI nanowire surface states that form the input and output states of the tri-junction. The spectrum of these surface states in the presence of an aligned external magnetic field is well described by~\cite{Rosenberg10,Cook11}
\begin{equation} \label{eq:spectrum}
    E(j, k) = \pm \hbar v_\Dirac \sqrt{k^2 + (2\pi j)^2 / P^2},
\end{equation}
with $v_\Dirac$ the Dirac velocity of the surface state cone, $k$ the wave vector along the direction of the wire, and $2\pi j/P$ the generalized transverse wave vector that contains contributions of the transverse orbital motion, the nontrivial Berry phase (i.e., equal to $\pi$), and the flux enclosed by the perimeter $P$ of the nanowire cross section.

\begin{figure*}[htb]
	\centering
	\includegraphics[width=.4435\linewidth]{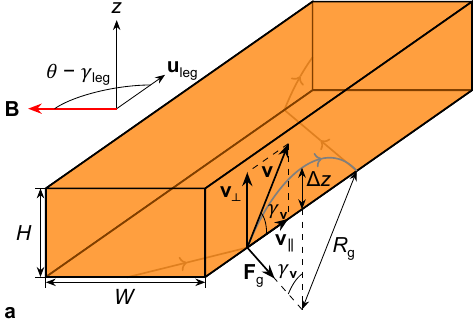}
	\hspace{.0\linewidth}
	\includegraphics[width=.5365\linewidth]{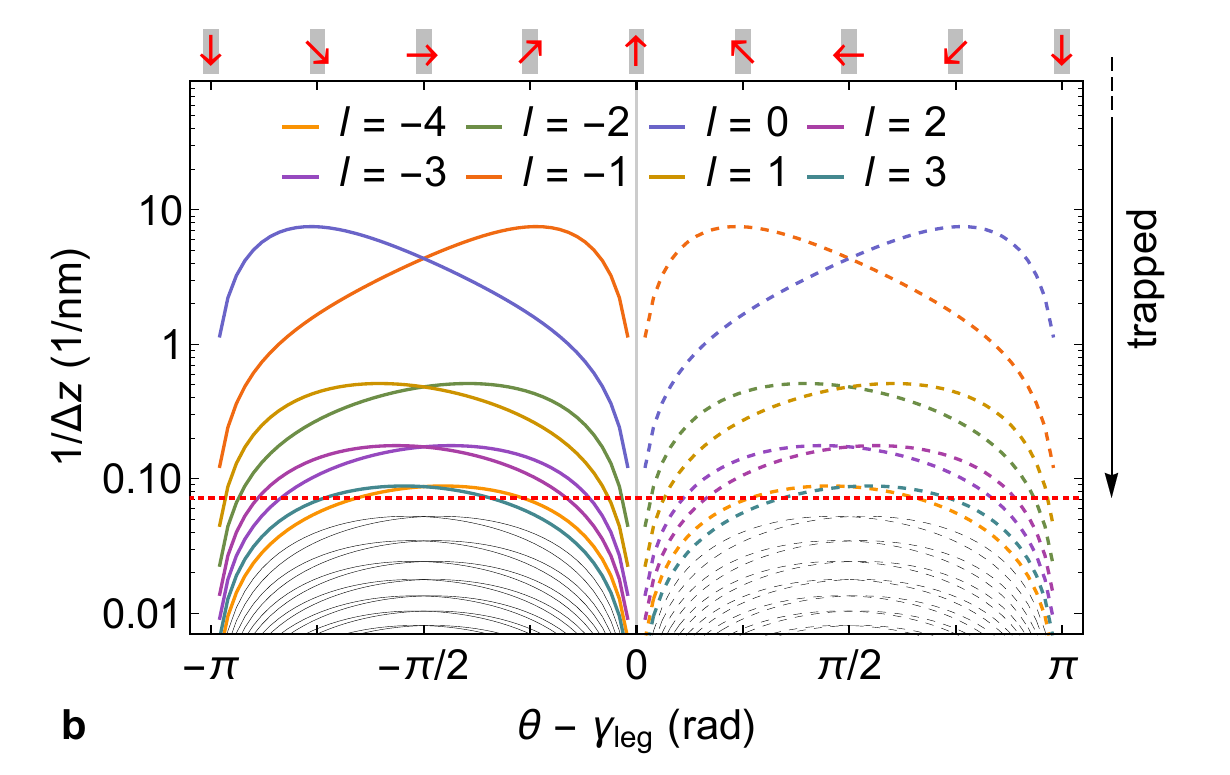}
	\caption{\textbf{Semiclassical model of the steering effect.}
		(\textbf{a}) The semiclassical trajectory of a 3D~TI nanoribbon surface state in the presence of a magnetic field that is not aligned with the ribbon. The maximal transverse distance $\Delta z$ that the surface state can travel on the side surface is indicated. In this case, the charge carrier cannot reach the top surface and is effectively trapped on the bottom surface.
		(\textbf{b}) The inverse transverse distance that the surface states of the different subbands [see Equation~\eqref{eq:spectrum}] are able to travel against the Lorentz force of a nonaligned in-plane magnetic field on the side surface of a nanoribbon is shown as a function of the angle $\theta-\gamma_\mathrm{leg}$ between the magnetic field and the ribbon orientation. Depending on this angle (and considering $v_\parallel > 0$), the direction of the force points towards the top (solid lines) or bottom (dashed lines) surface. The subband states that cannot traverse the side surfaces against the Lorentz force are indicated in color and the transverse-mode index $l$ is specified. A 150~nm-wide and 14~nm-high nanoribbon (based on the sample dimensions) with $E_\Fermi = 86\,\textnormal{meV}$ and $v_\Fermi = 5.5\times10^5\,\textnormal{m/s}$ (reasonable assumptions, corresponding to a 2D charge density of approximately $5\times10^{11}\,\textnormal{cm}^{-2}$), and a magnetic field strength of $0.5\,\textnormal{T}$ have been considered. The inverse wire thickness is indicated by the red dotted line.
	}
	\label{fig:Deltaz}
\end{figure*}

The crucial aspect for the steering effect is the orbital effect due to the nonaligned magnetic field component. It gives rise to a Lorentz force on the side facets of the nanoribbon that traps certain surface-state transport modes on the top or the bottom surface of the legs of the tri-junction, depending on the relative orientation of the in-plane magnetic field with respect to that leg. This type of trapping can be understood by considering the semiclassical trajectory of a surface-state charge carrier (see Fig.~\ref{fig:Deltaz}a). On the top and bottom surfaces of the nanowire, the trajectory is not affected, but a circular motion is induced on the side facets when the magnetic field is not perfectly aligned with the leg. The gyroradius $R_\mathrm{g}$ of that circular motion is given by
\begin{equation} \label{eq:gyroradius}
    R_\mathrm{g} = \left|E_\textsc{f} / (e B_\perp v_\textsc{f}) \right|,
\end{equation}
with $E_\textsc{F}$ the Fermi level energy (with respect to the Dirac point energy of the surface state cone), $e$ the elementary charge, $B_\perp = |\vecB|\cos(\theta-\gamma_\mathrm{leg})$ the component of the magnetic field perpendicular to the direction of the ribbon (with $\theta$ and $\gamma_\mathrm{leg}$ the orientation angles of the in-plane magnetic field and the ribbon, respectively). Based on this gyroradius, the height of the ribbon (approximately $14\,\textnormal{nm}$), and the velocity vector of a specific 3D~TI surface state on the side facet, we can estimate whether the surface state is able to traverse the side facet against the direction of the trapping (Lorentz) force. If this is not the case, the charge carrier is effectively trapped on the top or bottom surface of the nanoribbon while it moves towards (or away from) the junction, as the direction of the trapping force is the same on the two side facets of the ribbon. This direction is given by the sign of $\sin(\theta - \gamma_\mathrm{leg}) v_\parallel$, with $v_\parallel$ the velocity component of the surface state along the direction of the ribbon. The force points towards the top (bottom) surface when this sign is negative (positive).

The trapping effect can effectively block the transmission across a tri-junction to one of the output legs when the input channel and corresponding output channel on one of the output legs are trapped on opposite surfaces. Based on this trapping effect, we can construct a qualitative transmission model by assuming that the transmission across the tri-junction, which is otherwise expected to be left-right symmetric from general symmetry considerations, is suppressed when the incoming surface state and the corresponding output state are trapped on opposite surfaces (see Supplementary Note~\nolink{\ref{sec:details_model}} for a derivation of the model). The overall transmission is then obtained by summing over all the incoming surface-state channels of the input leg of the tri-junction, considering the geometry of the leg, the Fermi level energy for the 3D~TI surface states, and the external magnetic field.
The steering ratio obtained from the qualitative transmission model is compared to the experimentally obtained steering ratio profile in Fig.~\ref{fig:T-junction-B-constant} and they are in good qualitative agreement, both regarding the angular dependence as well as the field strength dependence, with the steering ratio profile becoming more pronounced as the magnetic field strength and corresponding trapping force increase.

\section*{Discussion} \label{sec:discussion}
The steering effect across a tri-junction due to trapped surface states on the top or bottom surface of a 3D~TI nanoribbon in the presence of an external in-plane magnetic field has been explained by considering semiclassical trajectories for the charge carriers. However, the phenomenology can also be confirmed with quantum transport simulations, considering a tight-binding model for 3D~TIs and making use of the software package Kwant~\cite{Groth14}. This analysis is presented in Supplementary Note~\nolink{\ref{sec:single-channel-limit}}. Further note that, as the topological surface states are spin-momentum-locked, the steering effect would naturally induce a steering of spin current as well, which could be considered for spin filter applications~\cite{Bellucci08}.

The in-plane magnetic field is not the only possible source of left-right symmetry breaking for the current in our experimental setup. An overview of the alternative symmetry breaking mechanisms is presented in Supplementary Note~\nolink{\ref{sec:alt_mechanisms}}, and it is discussed in detail what is their expected steering ratio profile and why they cannot be responsible for the observed $\pi$-periodic steering ratio profile.

While the steering ratio profiles obtained experimentally and from the transmission model are in good qualitative agreement, their amplitudes differ by a factor of around $25$. This discrepancy can be attributed to the contribution of bulk states to the total current in the experiment, something that is not taken into account in the transmission model. The bulk state properties are only weakly affected by the orbital effect of the magnetic field and are not subject to the trapping effect that affects the topological surface states. Hence, the scaling factor of $25$ is expected to reflect the ratio of bulk versus surface state current in the current measurements. While we cannot disentangle the bulk and surface state contributions in our experimental setup, a current contribution from the bulk that is up to one or even two orders of magnitude larger than the surface state contribution is expected, from the high charge carrier contribution that is obtained from the Hall bar characterization (see Methods section). Furthermore, considering that we have obtained the steering ratio profile from the transmission model assuming a surface state charge density of $\sim 1.5 \times 10^{12}\,\textnormal{cm}^{-2}$, this would imply a total charge density that is 25 times larger.

An important observation in support of our transmission model is the temperature dependence of the steering pattern. This crossover temperature, above which the steering pattern disappears, is in reasonable agreement with the temperature at which the phase-coherence length becomes comparable to the perimeter around the cross section of the legs of the tri-junction (see Methods section).
Hence, the temperature dependence of the individual currents signifies the importance of 3D~TI surface states retaining their phase coherence along the complete perimeter of the ribbon, similar to the magnetotransport pattern of flux quantum-periodic magnetoconductance oscillations appearing in straight 3D~TI nanoribbons at low temperatures~\cite{Arango16, Ziegler18, Rosenbach20}. In that case, the oscillations originate from the Aharonov-Bohm effect acting on the surface state subband spectrum, which requires phase coherence along the perimeter of the ribbon for the surface states to properly enclose the magnetic flux. In this case, however, it is the trapping effect that relies on the coherent propagation of the surface state quasiparticles along the perimeter of the ribbon such that the surface state solutions can become depleted on either the top or the bottom of the wire by the Lorentz force acting on the side facets. 

As the trapping effect is a direct consequence of the orbital effect on the surface-state charge carrier, we cannot immediately rule out the possibility that the steering effect is rooted in trivial surface states rather than topological surface states. For example, there could be surface states that originate from trivial bulk states due to bulk bending~\cite{Bianchi10}, which then get trapped by the external magnetic field. However, this scenario is less probable because of the following two reasons.
First, phase-coherence around the perimeter of the cross section, which is an important requirement for the trapping effect and is supported by the temperature dependence, as discussed in the previous paragraph, is much more difficult to realize with nontopological surface states, e.g., a conventional 2D electron gas, that are more sensitive to disorder and related localization effects.
Second, the trapping effect is expected to be more robust and pronounced for topological surface states as compared to trivial surface states because of spin-momentum locking. The trapping force flips sign when the velocity along the direction of the nanoribbon is reversed. For trivial surface states, such a change of the velocity can easily arise due to elastic scattering processes in the presence of disorder. For topological surface states, however, spin-momentum locking forbids direct backscattering and generally suppresses scattering events that flip the sign of the velocity.
For a more quantitative comparison, we performed quantum transport simulations of a T-junction in the single-channel regime with a topological surface state, a trivial surface state, and a bulk metallic state (see Supplementary Note~\nolink{\ref{sec:single-channel-limit}}). The resulting steering pattern agrees well with the trapping-based transmission model as well as the experimental data, and is robust against disorder, but only for the topological surface state.

In summary, we have observed an in-plane magnetic field-driven steering effect that breaks the left-right symmetry of the transmission of 3D~TI nanoribbon surface states across three-terminal junctions.
The effect can be attributed to the interplay of the phase-coherent topological surface states and the orbital effect on the side facets that causes these states to be trapped on the top or bottom surface of a nanoribbon, depending on the relative orientation of the ribbon and the magnetic field. This trapping effect can suppress the transmission to one of the two output legs of a 3D~TI nanoribbon-based tri-junction, which gives rise to a steering effect that can be understood from semiclassical considerations and quantum transport simulations. The steering effect is well described by a qualitative transmission model that provides good agreement with the theoretical findings and the experimentally obtained steering ratio profile of the electrical current.
The physical origin of the steering effect is corroborated by the temperature dependence of the steering ratio profile, which indicates the importance of the phase coherence of topological surface states around the full perimeter of the ribbon legs.
Our experimental and theoretical results reveal interesting magnetotransport properties of 3D~TI-based tri-junctions in the presence of an in-plane magnetic field that are relevant for their application in topological material-based quantum technologies.

\section*{Methods}
\subsection*{Growth \& fabrication} \label{sec:growth-and-fabrication}
In Fig.~\ref{fig:T-junction-SEM}a a schematic of the T-shaped sample layout is shown. For substrate preparation, first, a silicon (111) wafer was covered with a 6-nm-thick thermally-grown SiO$_2$ layer. Subsequently, a 25-nm-thick amorphous SiN$_x$ layer was deposited using plasma-enhanced chemical vapour deposition. The pattern for the subsequent selective-area growth was defined by electron beam lithography followed by reactive ion etching (CHF$_3$/O$_2$) and wet chemical etching using hydrofluoric acid. The 14-nm-thick topological insulator Bi$_2$Te$_3$ film was grown selectively by means of MBE on the Si(111) surface~\cite{Kampmeier16, Weyrich19, Koelzer20, Rosenbach20}. The epitaxial layer was capped in-situ by a $\sim$3-nm-thick AlO$_x$ layer~\cite{Lang11}. A scanning transmission electron micrograph of the cross section lamella of a 135-nm-wide nanoribbon as used in the T- and Y-junction prepared by focused ion beam milling is shown in Fig.~\ref{fig:T-junction-SEM}c. The epitaxial layers are crystallographically aligned with the substrate. The Ohmic contacts composed of a 5-nm-thick Ti layer and a 100-nm-thick Au layer were prepared by electron beam evaporation after development and removing the AlO$_x$ capping in the developed areas by argon sputtering. Finally, the device is capped with a HfO$_2$ layer by atomic layer deposition. In Fig.~\ref{fig:T-junction-SEM}b a scanning electron micrograph of an exemplary contacted T-junction is shown.

\subsection*{Material characterization} \label{sec:characterization}
Using a standard four-probe lock-in Hall setup in a variable temperature insert, Hall measurements were performed on devices of different sizes grown by means of MBE during the same run and published by Rosenbach \emph{et al}.~\cite{Rosenbach20}.
The authors find a total charge carrier concentration of $n_\mathrm{2D} = (6.8\text{ - }9.5) \times 10^{13}\,\textnormal{cm}^{-2}$ and a mobility of $\mu = (307\text{ - }374)\,\textnormal{cm}^{-2}$/Vs from analyzing the Hall data. 
Additionally, Shubnikov-de Haas oscillations are observed in a $500\,\mathrm{nm}$-wide Hall device. Based on these oscillations, the 2D sheet carrier concentration is found to be $n_\mathrm{SDH} = 5.3 \times 10^{11}\,\textnormal{cm}^{-2}$, at a mobility of $\mu = 1997\,\textnormal{cm}^{-2}$/Vs.
The latter values can be attributed to the topological surface states, which leads to an estimated Fermi level of $E_\Fermi = 86\,\mathrm{meV}$ with respect to the Dirac point.
Furthermore, a phase-coherence length of about $l_{\phi}(T < 1\,\textnormal{K}) \approx 240\,\textnormal{nm}$ is estimated from the magnetoresistance data in a nanoribbon. This value for $l_{\phi}$ is of the same order of magnitude as the perimeter of the cross section of the tri-junction legs for the device presented in this work.

\subsection*{Magnetotransport measurements} \label{sec:magnetotransport}
The magnetotransport measurements were carried out in a dilution refrigerator at a base temperature of $T = 25\,\textnormal{mK}$. The system is equipped with a $1\text{ - }1\text{ - }6\,\textnormal{T}$ vector magnet. For the electrical setup we refer to Fig.~\ref{fig:T-junction-SEM}b. In addition, every line is equipped with a set of filters adding a resistance of about $3.6\,\textnormal{k}\Omega$ each. In the experiment we use lock-in amplifiers at $f_\textnormal{LI} = 28.3\,\textnormal{Hz}$ and an operational amplifier-based voltage source to apply a bias voltage of $100\;\mu\textnormal{V}$ to the bottom terminal. The lock-in amplifiers are equipped with a current-to-voltage converter providing a virtual ground to the other two terminals. The voltage bias leads to a current $I_\total$ through the device input. The current then splits off into $I_\Ri$ and $I_\Le$, depending on the resistances of the two individual paths. Scanning the magnetic field along the different in-plane directions then yields the 2D current maps, as shown in Fig.~\ref{fig:T-junction-current-map} and in Supplementary Note~\nolink{\ref{sec:Y-junction}}.

\subsection*{Data analysis} \label{sec:dataanalysis}
In order to compare the experimental data with the theoretical model the magnetic field components of the steering ratio were transformed from Cartesian coordinates with point distances $\delta_{B_x} = \delta_{B_y} = 10\,\textnormal{mT}$ into polar coordinates and projected onto a rectangular grid with point distances $\delta_{B} = 6\,\textnormal{mT}$ and $\delta_{\phi} = 0.06\,\textnormal{rad}$ using linear interpolation.

\section*{Data availability}
The experimental data, the simulation source code and the simulated data that supports the findings of this study are available as source data files from the J\"{u}lich DATA repository (\url{https://doi.org/10.26165/JUELICH-DATA/CX10EO}).

%% file: supp_text.tex
\section{Temperature dependence} \label{sec:Y-junction}
Temperature is expected to affect the symmetry breaking and steering effect in two different ways.
First, the transmission across the junction will be averaged over a window around the Fermi energy proportional to $\kB T$. Around $T=1\,\textnormal{K}$ and below, this energy scale is much smaller than the other energy scales (in particular, the Fermi energy and the subband spacing of the 3D~TI nanoribbon surface states) that are relevant for the magnetotransport properties. Hence, this aspect is not expected to change the transport behavior significantly over the temperature ranges measured.
Second, the temperature has an impact on the phase-coherence length $l_\phi$ of the surface states, with a typical scaling relation of $l_\phi \propto T^{-1/2}$~\cite{Altshuler81a}. If the phase-coherence length becomes smaller than the perimeter of the wire $P$ ($l_\phi < P$) there is no coherent motion of surface state charge carriers around the perimeter of the nanoribbon and the surface states on the top and bottom surface become effectively decoupled. The trapping force due to the nonaligned magnetic field component cannot deplete the top or bottom surface in this case, which is essential for blocking transmission across the tri-junction and for the realization of the steering effect (see Supplementary Note~\ref{sec:details_model} for details). Hence, the steering ratio profile as a function of the magnetic field orientation angle is expected to vanish when $l_\phi > P$.

\begin{figure*}[htb]
    \centering
    \includegraphics[width=\linewidth]{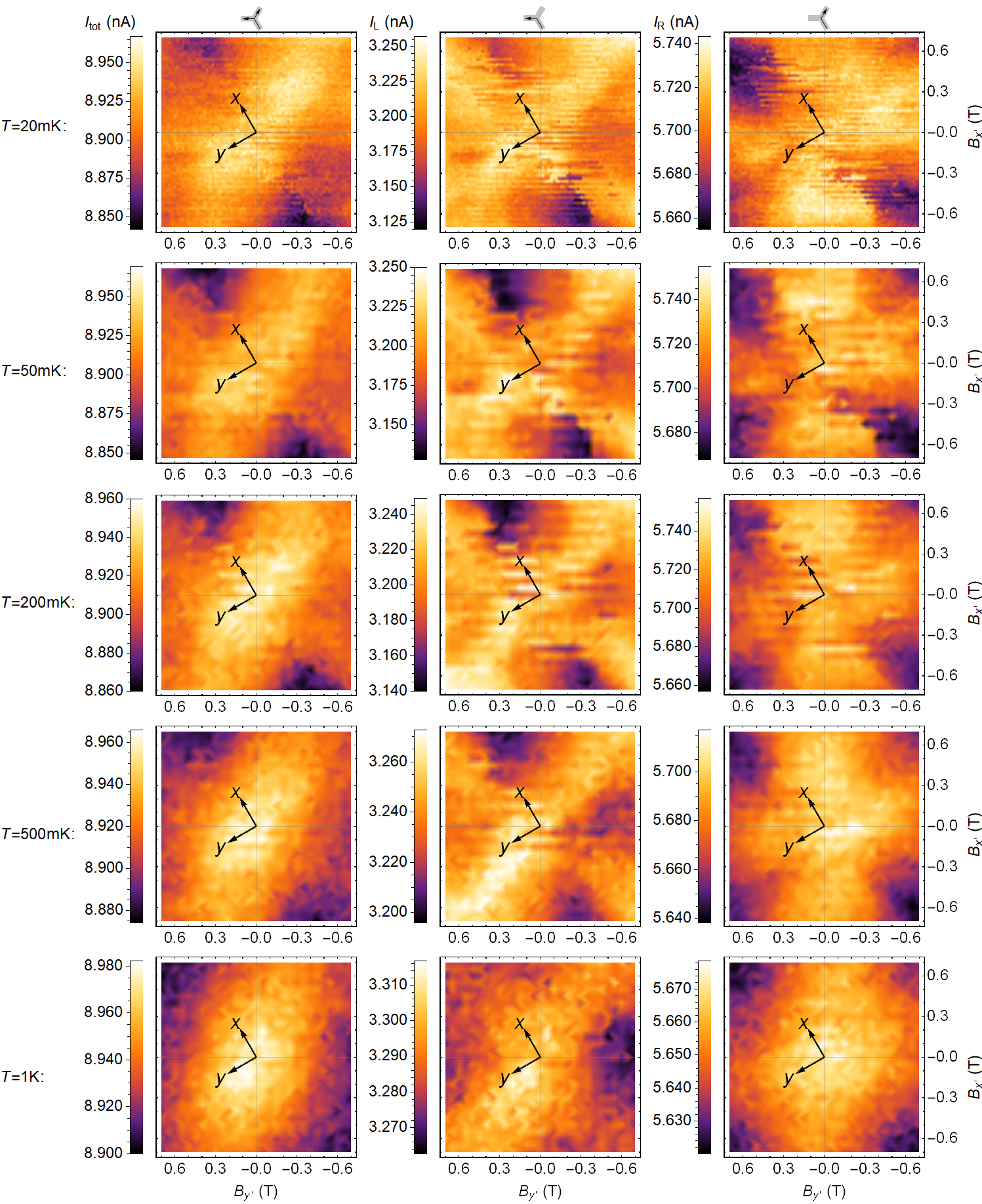}
    \caption{
        The current across a Y-shaped tri-junction as a function of the in-plane magnetic field components from the bottom leg to (left column) both output legs, and to (middle column) the left and (right column) the right output leg individually, measured at different temperatures.
    }
    \label{fig:Y-junction-T-dependence-raw}
\end{figure*}

The temperature dependence of the current as a function of the in-plane magnetic field components was characterized for a Y-shaped tri-junction with similar dimensions for the legs as the T-shaped tri-junction presented in the Main Text (see layout in Supplementary Fig.~\ref{fig:Y-junction-T-dependence-mean}a). Supplementary Figure~\ref{fig:Y-junction-T-dependence-raw} shows the temperature dependence of the total current (first column) and the individual steering currents (last two columns).
It is clearly visible that for an increase in the temperature (top to bottom) the emergent steering pattern starts to loose its contrast. This can be quantified by considering the standard deviation of the current with respect to the average over the different magnetic fields $\langle I \rangle_\mathbf{B}$ that have been applied (see Supplementary Figs.~\ref{fig:Y-junction-T-dependence-mean}b--d). This standard deviation is strongly decreasing at higher temperatures for the individual currents. The total current is not as isotropic as for the T-junction, however. The reason for this is that the T-junction allowed for a more symmetric transport setup. The outer contacts were chosen for all terminals due to their larger contact area and, hence, smaller contact resistance. Such a symmetric setup was not achieved in the Y-junction due to a high contact resistance, which is why the outer contact (see Supplementary Fig.~\ref{fig:Y-junction-T-dependence-mean}a) was selected for current injection and the inner contacts were used for extracting the current. The weak antilocalization affecting the different leads is therefore highly asymmetric, which superimposes on top of the junction-related transmission properties in the current signals and complicates the interpretation of the steering pattern.

\begin{figure*}[htb]
    \centering
    \includegraphics[width=0.85\linewidth]{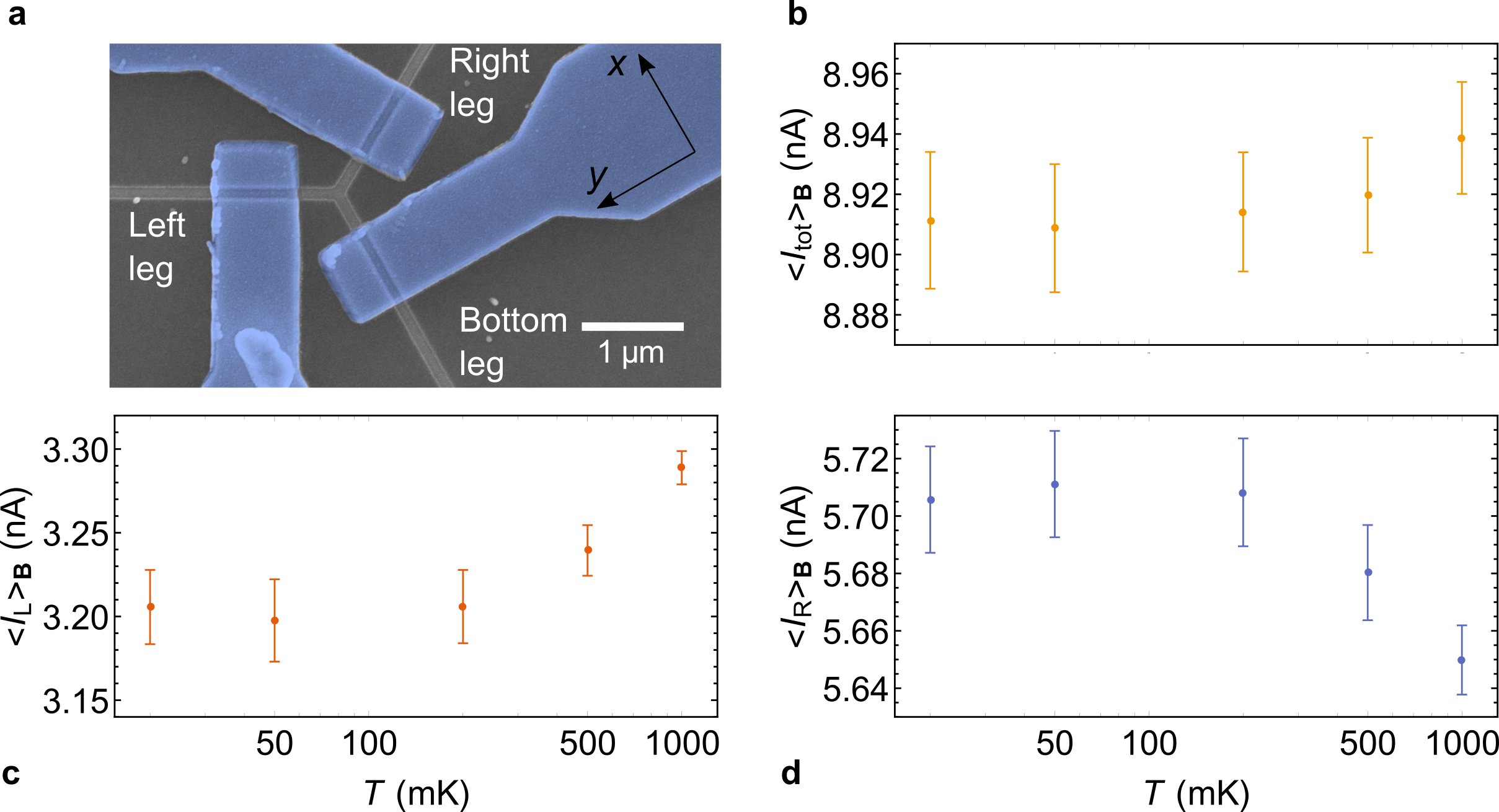}
    \caption{
        (\textbf{a}) False color scanning electron micrograph of the Y-junction device. The 3D~TI is colored in blue, the substrate in black, and the metallic contacts in gray. The voltage is provided at the outer terminal of the bottom leg and the current is measured at the inner contacts of the other two legs. The currents and reference frame for the external magnetic field components are indicated.
        (\textbf{b})--(\textbf{d}) The average over all applied in-plane magnetic field strengths and orientations (shown in Supplementary Fig.~\ref{fig:Y-junction-T-dependence-raw}) of the total current across a three-terminal junction Y-junction (in \textbf{b}), and of the individual currents to the left and right output legs (in \textbf{c} and \textbf{d}, respectively), is evaluated at different temperatures. The standard deviation with respect to this average $\Delta I$ (presented in Supplementary Fig.~\nolink{\ref{fig:Y-junction-T-dependence}} in the Main Text) is indicated by the error bars, ranging from $\langle I \rangle - \Delta I$ to $\langle I \rangle + \Delta I$.
    }
    \label{fig:Y-junction-T-dependence-mean}
\end{figure*}

\clearpage
\section{Details on transmission model for tri-junctions} \label{sec:details_model}
We establish the criteria for the transmission of 3D~TI nanowire (or nanoribbon) surface states across a tri-junction by considering the orbital effect of the external magnetic field on the surface state charge carriers in the different legs of the junction, as depicted in Fig.~\nolink{\ref{fig:Deltaz}}a in the Main Text.
The orbital effect of an in-plane magnetic field $\vecB \equiv (|\vecB| \cos \theta, |\vecB| \sin \theta, 0)$ yields the following Lorentz force on a charge carrier with velocity $\vecv$ and charge $-e$:
\begin{equation}
	\vecF = -e \vecv \times \vecB = -e |\vecB| \begin{pmatrix} - v_z \sin \theta \\ v_z \cos \theta \\ v_x \sin \theta - v_y \cos \theta \end{pmatrix}\, .
\end{equation}
The Lorentz force has no impact on the charge carriers when they are confined to a 2D surface parallel to the $(x, y)$-plane ($v_z = 0$), but it does affect the side facets of a nanowire when it is not perfectly aligned with the magnetic field. We proceed by considering a nanowire with rectangular cross section, top and bottom surfaces parallel to the $(x, y)$-plane, and side surfaces parallel to the plane spanned by unit vectors $\vecu_\side \equiv (\cos\gamma_\side, \sin\gamma_\side, 0)$ and $\vecu_z \equiv (0, 0, 1)$, with $\gamma_\side$ the in-plane orientation angle of the nanowire. On the top and bottom surfaces, the Lorentz force points perpendicular to the surface and does not affect the surface states. On the side surfaces, the velocity vector of a surface state can be written as $\vecv = v_\parallel \vecu_\side + v_\perp \vecu_z$ such that the resulting Lorentz force becomes:
\begin{equation}
	\vecF = -e|\vecB| \begin{pmatrix} - v_\perp \sin\theta \\ v_\perp \cos\theta \\ v_\parallel \sin(\theta - \gamma_\side) \end{pmatrix}.
\end{equation}
The force component that induces a circular motion on the side surface is given by $\vecF_\cycl \equiv \vecF - (\vecu_\side^\perp \cdot \vecF) \vecu_\side^\perp$, with $\vecu_\side^\perp \equiv (-\sin\gamma_\side, \cos\gamma_\side, 0)$ a unit vector perpendicular to the side surfaces of the nanowire, yielding:
\begin{equation}
	\vecF_\cycl = -e|\vecB| \sin(\theta - \gamma_\side) \begin{pmatrix} - v_\perp \cos\gamma_\side \\ - v_\perp \sin\gamma_\side \\ v_\parallel \end{pmatrix}.
\end{equation}
This component vanishes when the wire and magnetic field are aligned ($\theta = \gamma_\side$) and is maximal when the magnetic field is perpendicular to the side surface, initiating a clockwise or counterclockwise circular motion, depending on the relative orientation of the magnetic field and the nanowire, with gyroradius $R_\cycl$ given by:
\begin{equation}
	R_\cycl = \frac{m_\cycl |\vecv|^2}{|\vecF_\cycl|} = \left|\frac{E_\Fermi}{e |\vecB| \sin(\theta - \gamma_\side) v_\Dirac} \right|,
\end{equation}
with Fermi energy $E_\Fermi$ (relative to the Dirac point energy) and Dirac velocity $v_\Dirac$ of the 3D TI surface state spectrum, $E(\veck) = \pm \hbar v_\Dirac |\veck|$.
The surface state charge carriers generally accelerate towards the top or bottom surface when entering the side surface, depending on their initial velocity vector and the in-plane orientation of the magnetic field relative to the nanowire, according to the following rule:
\begin{equation} \label{eq:trapping_rule}
	\textnormal{force/acceleration towards the} \begin{cases} \textnormal{top} \\ \textnormal{bottom} \end{cases} \textnormal{ surface if } \sign[-\sin(\theta - \gamma_\side) v_\parallel] = \begin{cases} 1 \\ -1 \end{cases}.
\end{equation}
To estimate the effectiveness of the magnetic field in preventing the surface state charge carriers from traversing the side facet from top to bottom or vice versa, we consider the distance $\Delta z$ that a surface state charge carrier can travel in the transverse direction before its transverse velocity component is reversed (see Fig.~\nolink{\ref{fig:Deltaz}}a in Main Text):
\begin{equation} \label{eq:orbital_effect_sides}
	\Delta z = R_\cycl (1 - \cos\gamma_\vecv) = \left|\frac{E_\Fermi}{e |\vecB| \sin(\theta - \gamma_\side) v_\Dirac} \right| \lef 1 - \frac{v_\parallel}{v_\Dirac} \rig.
\end{equation}
If $\Delta z < H$, with $H$ the height of the side facet, the surface state charge carrier cannot traverse the side against the direction of the force (in the semiclassical picture) and is effectively trapped on the top or bottom surface (neglecting the extension up to $\Delta z$ on the sides). Even in the case that $R_\cycl \gg H$, several transverse modes with $v_\parallel \approx v_\Dirac$ can get trapped in this way.

Considering the spectrum of subbands of a 3D~TI nanowire in the presence of an aligned magnetic field [see Equation~\nolink{\eqref{eq:spectrum}} and explanation below in Main Text], there is a discrete set of allowed velocity vectors at the Fermi level with $E(j, k) = \pm \hbar v_\Dirac \sqrt{k^2 + (2 \pi j)^2/P^2} = E_\Fermi$, $v_\parallel = v_\Dirac k/\sqrt{k^2 + (2 \pi j)^2/P^2}$, and $v_\perp = \sqrt{v_\Dirac^2-v_\parallel^2}$. Correspondingly, we obtain a subband-dependent transverse extension on the side surfaces when considering the nonaligned component of the external magnetic field on the semiclassical trajectories, given by:
\begin{equation} \label{eq:Delta_z}
	\Delta z(j, k) = \left|\frac{E_\Fermi}{e |\vecB| \sin(\theta - \gamma_\side) v_\Dirac} \right| \lef 1 - \frac{|k|}{\sqrt{k^2 + (2 \pi j)^2/P^2}} \rig.
\end{equation}
In Fig.~\nolink{\ref{fig:Deltaz}}b in the Main Text, the inverse of this transverse distance is shown as a function of the angle between nanowire and (in-plane) magnetic field orientation for the different transverse modes labeled by integer $l = j - 1/2 - B_\parallel A/\Phi_0$. If $\Delta z < H$, the direction of the force determines the surface on which the surface state is trapped. Because the orientation angle of the different legs of the junction is different, the states with the identical corresponding quantum number can become trapped on opposite surfaces for certain magnetic field orientations. When this scenario applies, it can be expected that the transmission is suppressed. This was also confirmed with quantum transport simulations of a T-shaped three-terminal junction (see Supplementary Note~\ref{sec:single-channel-limit}).

Based on the magnetic field-induced trapping of the 3D~TI surface state charge carriers, we propose the following set of rules for the transmission across a three-terminal junction in the presence of an external magnetic field that is applied in the plane of the junction (also see Supplementary Fig.~\ref{fig:decision_tree} for the corresponding decision tree):
\begin{itemize}
	\item An input state (moving towards the tri-junction) with energy $E$ and transverse-mode index $l$ can only exit as an output state (moving away from the junction) with identical energy\footnote{We are considering elastic scattering across the junction.} and transverse-mode index (quantum number) across the junction. We refer to these output states as \emph{valid} output states.
	\item If there are multiple valid output states, the transmission across the junction to these different output states has equal probability.
	\item If the input state is trapped on the bottom or top surface of the input leg, and there are valid output states that are trapped on the opposite surface of the output leg, transmission across the junction to those output states is suppressed (i.e., there is no transmission to that output state).
\end{itemize}
This set of rules takes into account the impact of the aligned and nonaligned components of the external magnetic field on the surface states of the different legs and qualitatively describes the results that follow from quantum transport simulations.
Note that the conservation of transverse-mode index $l$, while not being a generally valid assumption, allows for a straightforward evaluation of the transmission coefficients. This simplification affects the precise value of the obtained steering ratios, which are difficult to compare directly with experiment because of the (large) nonsteering bulk contribution. The shape of the profile as a function of in-plane magnetic field orientation, however, is not affected and is the crucial and distinct property. Further note that the nonaligned only determines trapping in this model and not the subband spectrum itself. While the detailed tight-binding simulations, presented in Supplementary Note~\ref{sec:single-channel-limit} below, show that the spectrum is also affected, this aspect does not affect the steering ratio profile qualitatively.
This transmission model is considered to obtain Fig.~\nolink{\ref{fig:T-junction-B-constant}}b in the Main Text, assuming $v_\Fermi = 5.5\times10^5\,\textnormal{m/s}$ and $E_\Fermi = 86\,\textnormal{meV}$, which are reasonable assumptions for the 3D~TI material of the sample~\cite{Rosenbach20S} and correspond to a topological surface state charge density of approximately $5\times10^{11}\,\textnormal{cm}^{-2}$, and a $150 \times 14\,\textnormal{nm}^2$ rectangular cross section, based on the dimensions of the samples.

\begin{figure}[hbt]
	\centering
	\includegraphics[width=0.9\linewidth]{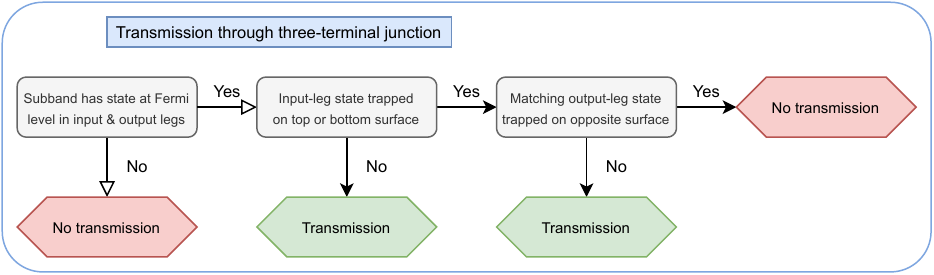}
	\caption{
		A decision tree with the rules of transmission across a tri-junction, according to the transmission model for in-plane magnetic field-driven current steering.
	}
	\label{fig:decision_tree}
\end{figure}

\clearpage
\section{Quantum transport simulations} \label{sec:single-channel-limit}
\subsection{3D~TI-based T-junction} \label{subsec:3DTI-T}
In this subsection, we present the confirmation of the trapping effect and the transmission model presented in Supplementary Note~\ref{sec:details_model} above, which qualitatively describe in-plane magnetic field-driven symmetry breaking and current steering in three-terminal junctions. For this, we performed quantum transport simulations in the single-channel regime, using the simulation package Kwant~\cite{Groth14S}. This approach has already been applied to study the in-plane magnetic field dependence of kinks and Y-junctions in Ref.~\cite{Moors18S}. Here, we present the analysis of a T-junction. We consider the following three-dimensional effective continuum Hamiltonian for 3D~TIs:
\begin{equation}
\begin{split}
    \mathcal{H} (\veck) &\equiv \epsilon(\veck) + \tau_z M(\veck) + \tau_x A_\perp (\sigma_x k_x + \sigma_y k_y) + \tau_x \sigma_z  A_z k_z, \\
		\epsilon(\veck) &\equiv C_0 - C_\perp (k_x^2 + k_y^2) - C_z k_z^2, \quad
			M(\veck) \equiv M_0 - M_\perp(k_x^2 + k_y^2) - M_z k_z^2.
\end{split}
\end{equation}
This Hamiltonian accurately describes a gapped bulk spectrum and a gapless surface state Dirac cone for the proper choice of parameters.
This Hamiltonian is then discretized on an artificial cubic lattice with lattice constant equal to 1~nm for the construction of T-shaped three-terminal junction.

As for the analytical cylindrical-nanowire model, the surface state spectrum of a nanowire with arbitrary cross section is subband-quantized with a confinement gap opening up at the Dirac point energy and a doubly degenerate spectrum (see Supplementary Fig.~\ref{fig:SR_T_kwant}a).
The aligned component of the magnetic field shifts the quantized transverse wave vectors through an Aharonov-Bohm phase, which lifts the double degeneracy in general, while the nonaligned external magnetic field shifts the Dirac cone in reciprocal space along the transport direction and flattens the subbands towards the Landau level regime (see Supplementary Fig.~\ref{fig:SR_T_kwant}b).

The steering ratio is extracted from the scattering matrix that is calculated with Kwant in the single-channel regime for different magnetic field strengths and orientations, as shown in Supplementary Fig.~\ref{fig:SR_T_kwant}c. The steering pattern that emerges is similar to what was observed for the kink and Y-junction nanostructures in Ref.~\cite{Moors18S} and similar to the profile obtained from the qualitative transmission model presented in Supplementary Note~\ref{sec:details_model}.

The trapping effect that was explained in Supplementary Note~\ref{sec:details_model}, based on the Lorentz force in the semiclassical picture, can also be confirmed by resolving the wave function density for the single input and output channel in each leg of the junction. In the presence of a nonaligned component of the in-plane magnetic field, the modes are confined to the top or bottom surface, depending on the relative orientation of the leg and the in-plane magnetic field. An overview of the density in the different legs with an external in-plane magnetic field along one of the four diagonals is presented in Supplementary Fig.~\ref{fig:density_T_kwant}. The density profile agrees with what is expected from the direction of the trapping force in Supplementary Equation~\eqref{eq:trapping_rule}.

\begin{figure}[hb]
	\centering
	\includegraphics[width=.6\linewidth]{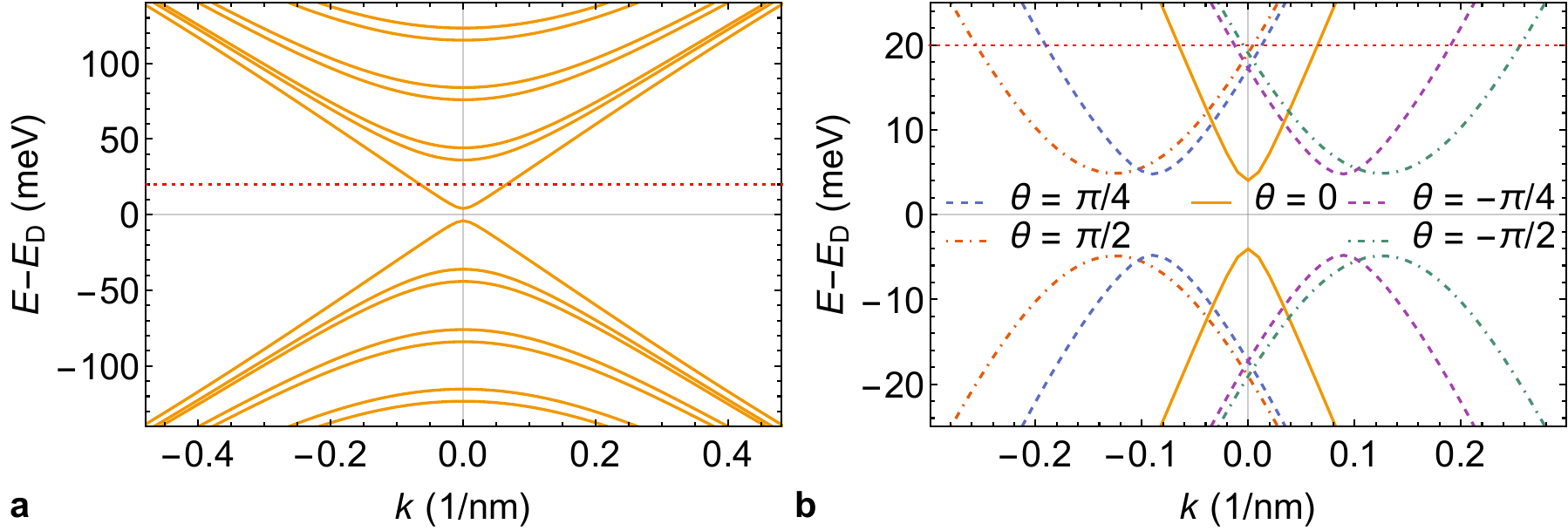}
	\includegraphics[width=.35\linewidth]{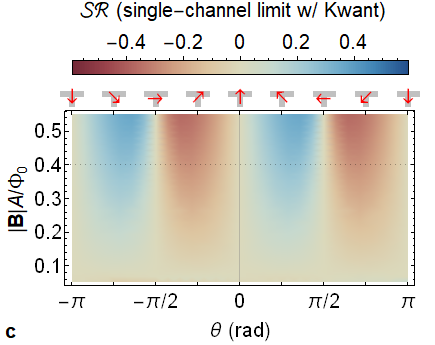}
	\caption{
	    (\textbf{a}) The subband-quantized Dirac cone spectrum of 3D TI nanowire surface states with an external magnetic field along the wire direction.
		(\textbf{b}) The surface state energy spectrum near the Dirac point energy $E_\Dirac$ for different in-plane magnetic field orientation angles.
		(\textbf{c}) The steering ratio of a T-junction as a function of in-plane magnetic field orientation angle $\theta$ and magnetic field strength $|\vecB|$ in units of flux quanta ($\Phi_0 \equiv h/e$) piercing the nanowire cross section.
		The nanowire cross section is equal to $A = 10 \times 10$~nm${}^2$, and the 3D TI model parameters are given by $A_\perp = A_z = 3$~eV$\cdot$\AA, $M_0 = 0.3$~eV, $M_\perp = M_z = 15$~eV$\cdot$\AA${}^2$, $C_0 = C_\perp = C_z = 0$. The magnetic field strength in \textbf{a}--\textbf{b} is equal to $0.8 \, \Phi_0/(2A)$ (black dotted line in \textbf{c}), and the energy at which the steering ratio is evaluated in \textbf{c} is 20~meV above the Dirac point energy (red dotted line in \textbf{a}--\textbf{b}).
	}
	\label{fig:SR_T_kwant}
\end{figure}

\begin{figure}
	\centering
	\includegraphics[width=.75\linewidth]{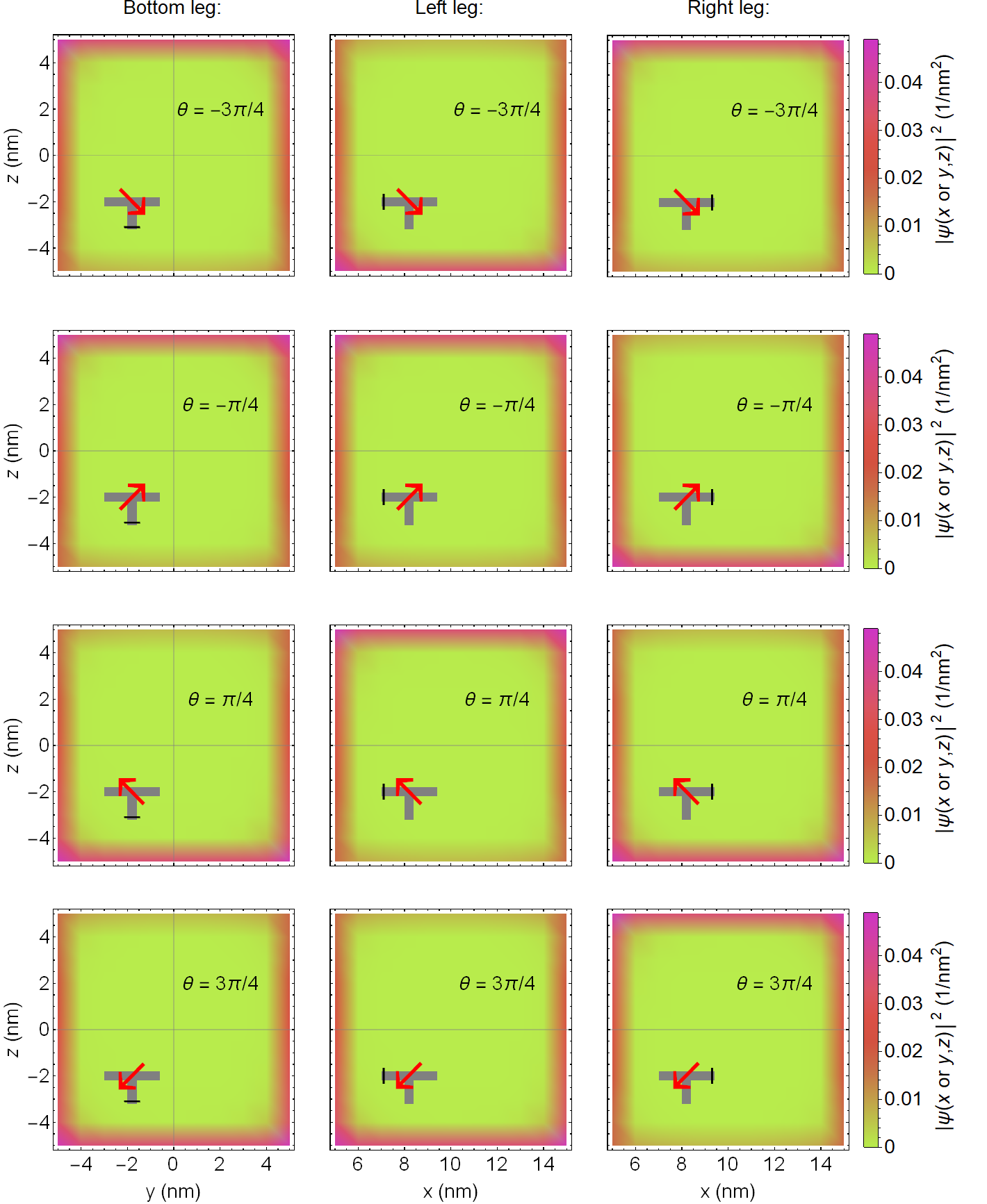}
	\caption{
	    The wave function density the single channel at 20~meV above the Dirac point energy in the three different legs of a T-junction, with the geometry of the legs, the 3D TI model Hamiltonian parameters, and the magnetic field strength the same as in Supplementary Figs.~\ref{fig:SR_T_kwant}a--b. The density in the different legs is presented in the three columns (for bottom, left, and right leg, respectively) for the four different diagonal in-plane magnetic field orientations in the different rows (indicated by the red arrow in the inset).
	    The incoming mode is shown for the bottom leg, and the exit modes for the left and right legs.
	}
	\label{fig:density_T_kwant}
\end{figure}

\subsection{Comparison between topological surface-state, trivial surface-state, and bulk channels} \label{subsec:comparison}
In this subsection, we compare the steering effect of the topological surface-state channel of a 3D~TI nanowire with a trivial two dimensional electron gas (2DEG) surface-state channel and a bulk channel. For the trivial 2DEG and the bulk channel, we consider the following single-band continuum models that are discretized on the same artificial cubic lattice as the one considered for the 3D~TI-based junction in the previous subsection:
\begin{equation}
\begin{aligned}
    \mathcal{H}_\text{metal} (\veck) &\equiv \frac{\hbar^2}{2 m_e^\ast} (k_x^2 + k_y^2 + k_z^2), \\
    \mathcal{H}_\text{2DEG} (\mathbf{r}, \veck) &\equiv \frac{\hbar^2}{2 m_e^\ast} (k_x^2 + k_y^2 + k_z^2) + V_\text{2DEG}(\mathbf{r}),
\end{aligned}
\end{equation}
with effective mass $m_e^\ast = 0.58\,m_e$ (appropriate for the bulk conduction band of Bi$_2$Te$_3$~\cite{Lee1987}), $V_\text{2DEG}(\mathbf{r}) = V_\text{2DEG}$ in the bulk and $V_\text{2DEG} = 0$ on the surface. For the comparison, we consider a 2DEG with a surface thickness of a single layer of lattice sites (note that the lattice constant equals $1\,\textnormal{nm}$) and a bulk potential equal to $V_\text{2DEG} = 0.3\,\textnormal{eV}$, which is the same as the energy distance between the Dirac point and the bottom of the conduction band in the 3D~TI continuum model that we have considered in the previous subsection.

\begin{figure}[hb]
	\centering
	\includegraphics[width=.99\linewidth]{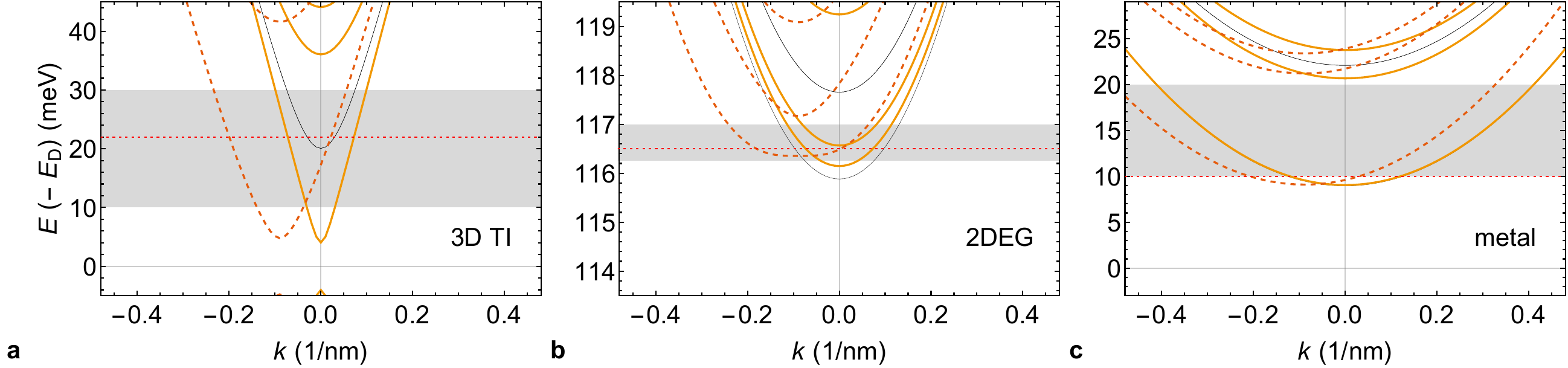}
	\caption{
	    (\textbf{a})-(\textbf{c}) The subband-quantized spectrum of (\textbf{a}) a 3D TI nanowire, (\textbf{b}) a bulk-insulating nanowire with trivial 2DEG states, and (\textbf{c}) a metal nanowire is shown with yellow solid and orange dashed lines, corresponding to a setup with an external magnetic field aligned with the wire, and under an angle of 45 degrees, respectively.
	    The nanowire cross section is equal to $A = 10 \times 10$~nm${}^2$, and the 3D~TI model parameters are the same as in Supplementary Fig.~\ref{fig:SR_T_kwant}. The parametrization of the metal and the 2DEG are specified in the text in Supplementary Note~\ref{subsec:comparison}.
	    A magnetic field strength equal to $0.8 \, \Phi_0/(2A)$ is considered.
	    The nanowire spectrum without an external magnetic field is also presented, with thin black lines.
	    The energy of the states whose wave function density is shown in Supplementary Fig.~\ref{fig:wave-function-density_comparison} is indicated by a red dotted line, and the energy range of Supplementary Fig.~\ref{fig:transport_comparison} by a gray zone.
	}
	\label{fig:spectrum_comparison}
\end{figure}

The electronic band structure near the single-channel regime is displayed for the different systems in Supplementary Fig.~\ref{fig:spectrum_comparison}.
The wave function densities over the cross section in the presence of a nonaligned magnetic field are presented in Supplementary Fig.~\ref{fig:wave-function-density_comparison}, similarly as in Supplementary Fig.~\ref{fig:density_T_kwant}. It is clear that both the topological and 2DEG surface states are subject to the magnetic field-induced trapping effect due to the orbital effect of the nonaligned field component, while the metal bulk state is barely affected by the external magnetic field.
The total transmission and the steering ratio across a T-junction, constructed with the three different tight-binding models, are presented as a function of energy and a rotating in-plane magnetic field in Supplementary Fig.~\ref{fig:transport_comparison}.
While the topological and trivial surface states are both subject to the trapping effect in the junction legs, it is not clearly reflected in the steering pattern of the trivial 2DEG channel. There is strong steering, but it does not align with the expected pattern based on trapping on identical or opposite side facets of the wire. Furthermore, the total transmission is also strongly dependent on the magnetic field orientation and the energy of the input channel.
In contrast, the 3D~TI system shows near-perfect total transmission that is only weakly modulated by the magnetic field orientation as well as a clean steering pattern that agrees well with the effective trapping-based transmission model and experimentally measured steering ratios.
The main qualitative difference between the 3D~TI and bulk-insulating system with trivial 2DEG surface states is spin-momentum locking of the topological surface states, which appears to provide the required robustness for trapping-based transmission and steering across a tri-junction by suppressing (back)scattering processes. This is further supported by the results showing the impact of disorder in the following subsection.

\begin{figure}[bt]
	\centering
	\includegraphics[width=.75\linewidth]{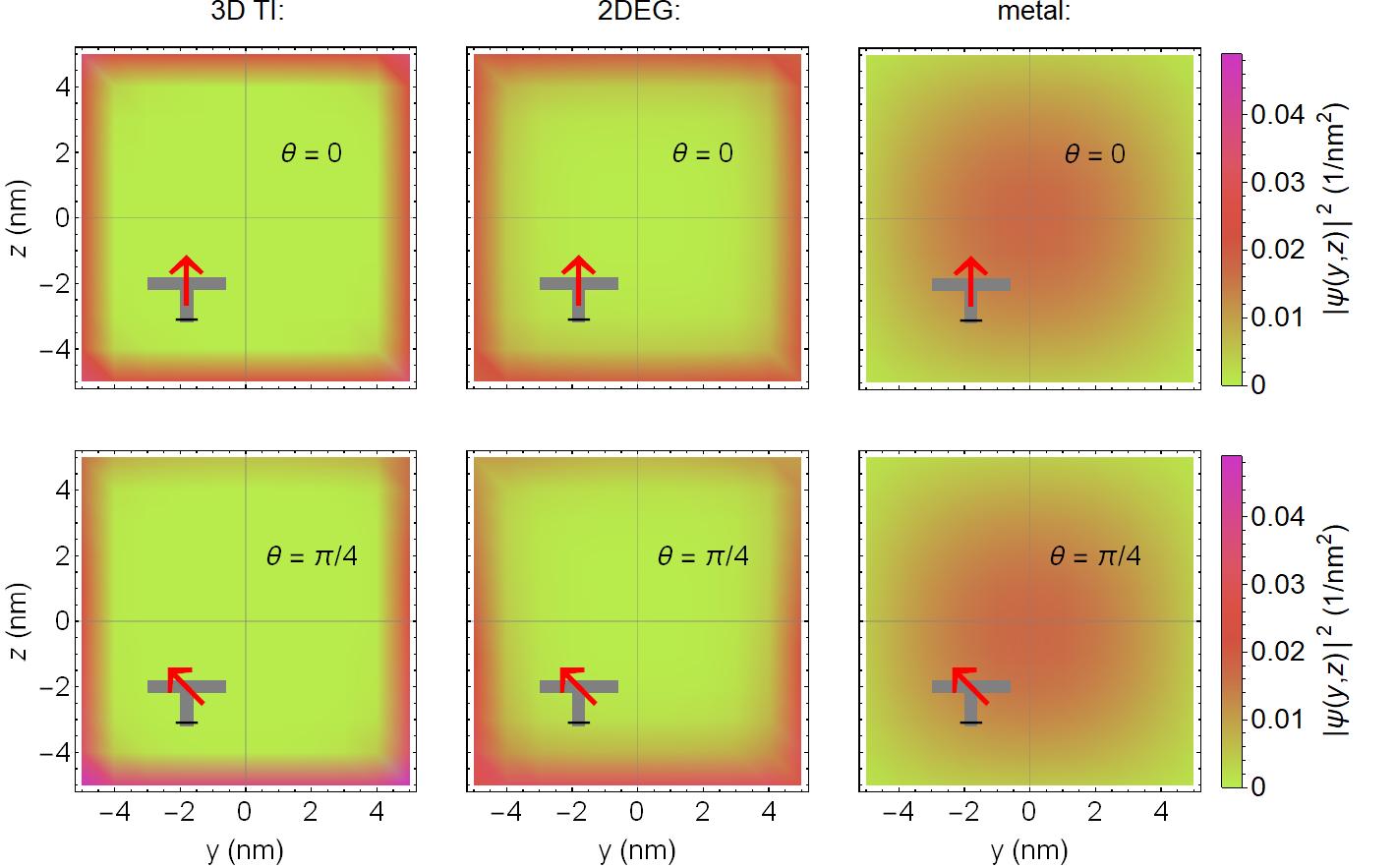}
	\caption{
	    The wave function density of the lowest-energy channel of the nanowire systems (or bottom leg of a T-junction) presented and compared in Supplementary Fig.~\ref{fig:spectrum_comparison} in the presence of an external magnetic field (top row) aligned with the nanowire and (bottom row) under an angle of 45 degrees with respect to the nanowire direction.
	    A magnetic field strength equal to $0.8 \, \Phi_0/(2A)$ is considered.
	}
	\label{fig:wave-function-density_comparison}
\end{figure}

\begin{figure}[bt]
	\centering
	\includegraphics[width=\linewidth]{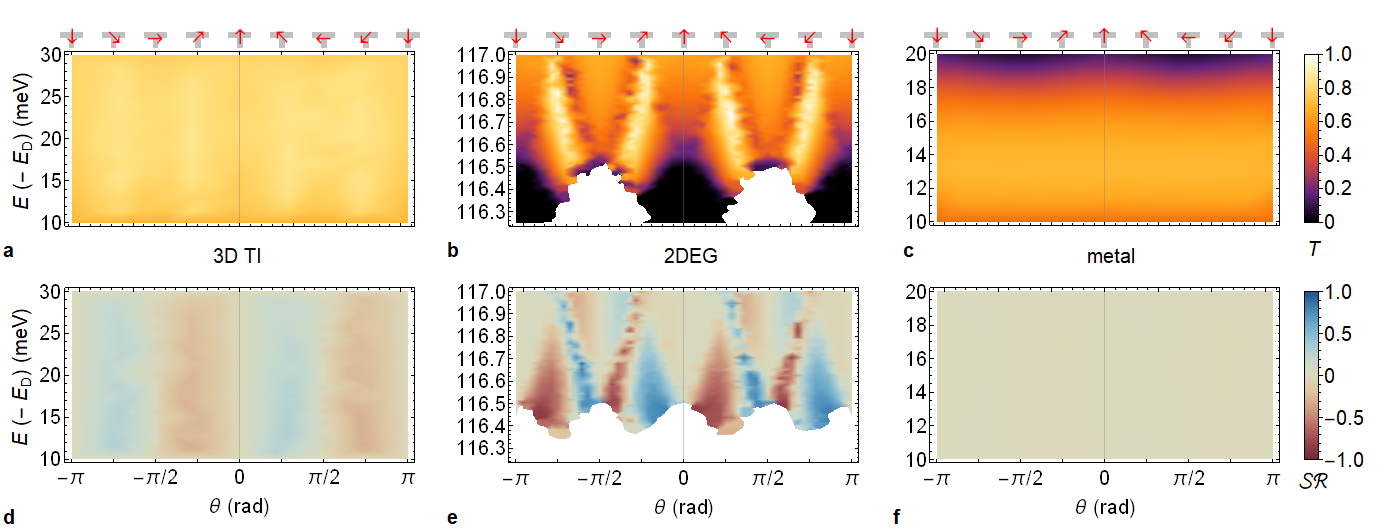}
	\caption{
	    (\textbf{a})-(\textbf{c}) The total transmission probability $T$ and (\textbf{d})-(\textbf{f}) the steering ratio over a T-junction (with bottom leg as input) as a function of energy $E$ and the external magnetic field orientation angle $\theta$, considering junction legs of (\textbf{a}), (\textbf{d}) a 3D~TI nanowire with topological surface states, (\textbf{b}), (\textbf{e}) a bulk-insulating nanowire with 2DEG surface states (note the transport regions in which there are no input or output channels), and (\textbf{c}), (\textbf{f}) a metal nanowire, all of them in the single-channel regime (see spectrum in Supplementary Fig.~\ref{fig:spectrum_comparison}), considering a field strength equal to $0.8 \, \Phi_0/(2A)$.
	}
	\label{fig:transport_comparison}
\end{figure}

\subsubsection{Impact of disorder} \label{subsubsec:disorder}
We model disorder by adding a random onsite potential to each lattice site of the tight-binding model with lattice constant equal to 1~nm. The onsite potential is picked independently for each site from a uniform interval $[-V_\text{dis}/2, +V_\text{dis}/2]$, considering $V_\text{dis} = 0.05\,\textnormal{eV}$, and a disorder average is considered over ten simulations with different disorder samplings.
The average and standard deviation of the total transmission and steering ratio obtained in this way are presented in Supplementary Fig.~\ref{fig:disorder_comparison}, together with the results of the pristine system without disorder. There is a clear qualitative difference between the result for the topological surface state of the 3D~TI and the two other systems. Both the total transmission and the steering pattern of the topological surface state are barely affected by disorder, whereas the patterns are strongly affected by disorder for the bulk channel of the metal system and the trivial 2DEG surface state of the bulk-insulating system. In particular, the pronounced orientation angle-dependent pattern of the total transmission and steering ratio for the 2DEG gets heavily suppressed when introducing disorder in the system. Only the steering pattern of the 3D~TI that matches the expected pattern based on the trapping effect and agrees with the experimentally obtained pattern shows any robustness with respect to disorder.

\begin{figure}[bt]
	\centering
	\includegraphics[width=\linewidth]{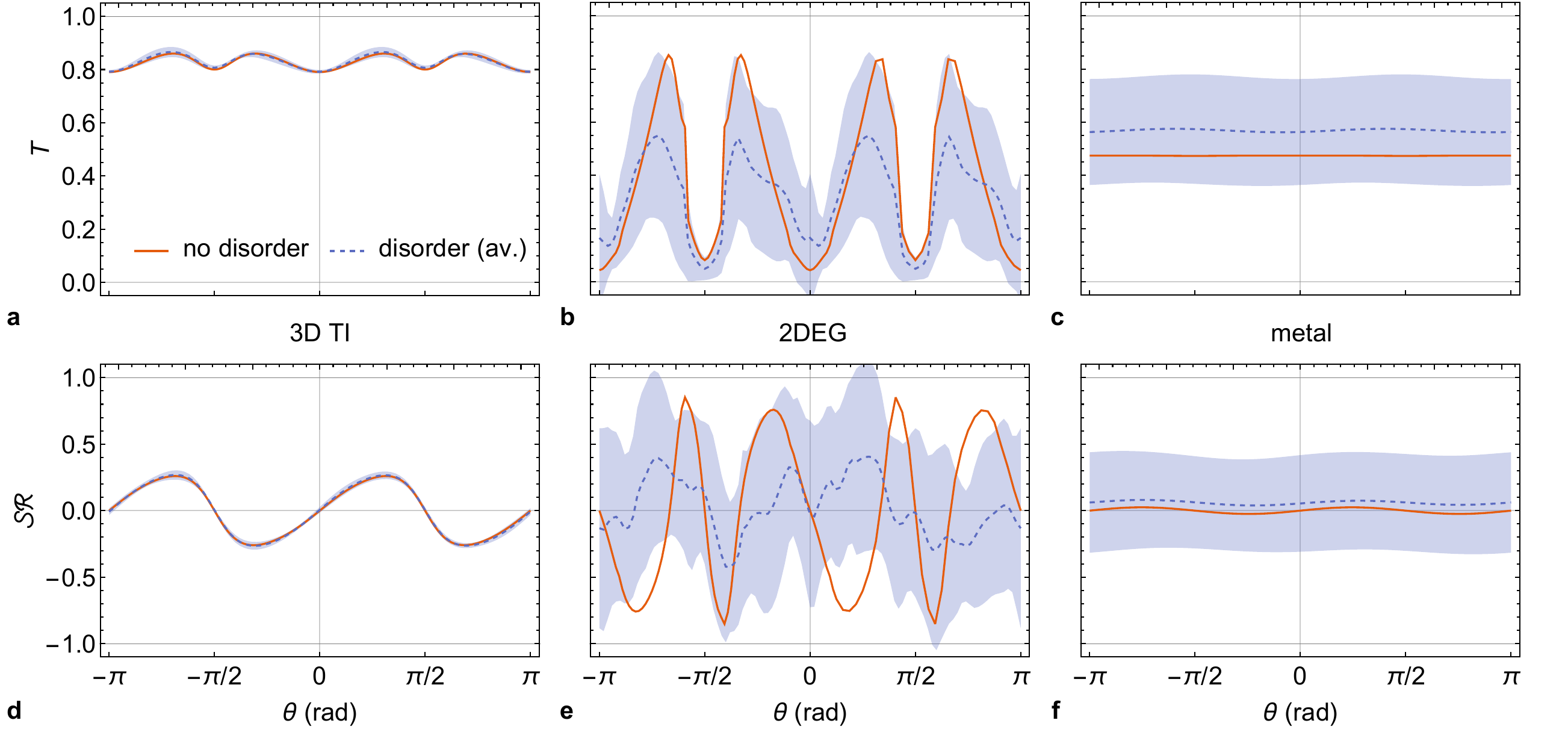}
	\caption{
	    (\textbf{a})-(\textbf{c}) The total transmission probability $T$ and (\textbf{d})-(\textbf{f}) the steering ratio over a pristine (solid lines) and disordered (dashed lines) T-junction (with bottom leg as input) as a function of the external magnetic field orientation angle $\theta$, considering junction legs of (\textbf{a}), (\textbf{d}) a 3D~TI nanowire with topological surface states, (\textbf{b}), (\textbf{e}) a bulk-insulating nanowire with 2DEG surface states, and (\textbf{c}), (\textbf{f}) a metal nanowire, all of them in the single-channel regime (see spectrum in Supplementary Fig.~\ref{fig:spectrum_comparison}), considering a field strength equal to $0.8 \, \Phi_0/(2A)$ and input-mode energy depicted by the red dotted line in Supplementary Fig.~\ref{fig:spectrum_comparison}.
	    The result for disorder is an average over ten different simulations with the same disorder strength (details in text) and the spread around the dashed curve indicates the standard deviation of the result.
	}
	\label{fig:disorder_comparison}
\end{figure}

\clearpage
\section{Alternative symmetry-breaking mechanisms} \label{sec:alt_mechanisms}

\subsection{Out-of-plane magnetic field component}
The orbital effect of an out-of-plane magnetic field $(0, 0, B_\perp)$ acting on an electron yields the following Lorentz force:
\begin{equation}
    \vecF_\perp = -e B_\perp \begin{pmatrix} v_y \\ -v_x \\ 0 \end{pmatrix}.
\end{equation}
The gyroradius $R_\cycl^\perp$ is then given by:
\begin{equation}
    R_\cycl^\perp = m_\cycl \sqrt{v_x^2 + v_y^2}/(e |B_\perp|),
\end{equation}
with $m_\cycl$ the energy-dependent cyclotron effective mass of the massless 3D TI surface states: $m_\cycl = (\hbar^2 / 2 \pi) (\partial S / \partial E) = E_\Fermi / v_\Dirac^2$, with $S(E) = \pi k(E)^2 = \pi E^2 / (\hbar ^2 v_\Dirac^2)$ the area in reciprocal space that is enclosed by the circular orbit.
If the external magnetic field has an approximate in-plane orientation and a $0.5$\,T field strength, and there is a $5^\circ$ misalignment between the plane of the magnetic field and that of the sample plane (the misalignment in our setup is at most a few degrees), the gyroradius for the maximal out-of-plane component ($0.5$\,T $\times \sin 5^\circ$) is approximately equal to $R_\cycl^\perp \approx 6.6\;\mu\textnormal{m}$. By comparing this length scale to the sample dimensions, the orbital effect of the out-of-plane magnetic field component can safely be neglected.

If the out-of-plane component would become relevant due to a large magnetic field strength and misalignment angle, the direction of the circular orbit would be determined by the sign of the out-of-plane component. With the angle $\theta = \theta_\mathrm{mis}$ of maximal misalignment of the external magnetic field (with $B_z > 0$, a steering ratio pattern proportional to $\SR \propto - \cos(\theta - \theta_\mathrm{mis})$ would be expected [considering the definition of the steering ratio in Equation~\nolink{\eqref{eq:steering_ratio}} in the Main Text]. This is a $2\pi$-periodic steering pattern that cannot be identified in the experimental current data, which is consistent with our estimation above.

\subsection{Zeeman coupling}
The transmission model presented in Supplementary Note~\ref{sec:details_model} only considers the orbital effect of the magnetic field, which in general breaks the degeneracy of the $l$ and $-l-1$ transverse modes. Zeeman coupling can break the degeneracy of two states with the same transverse-mode index and with opposite spin and momentum, however. Considering a $g$-factor of $g \approx 30$~\cite{Cook11S}, we obtain a maximal energy splitting $g \mu_\Bohr |\vecB|$ yielding $\sim1.7\,\textnormal{meV/T}$ between states with opposite spin and external magnetic field aligned with that spin, whereas the orbital effect induces a splitting $4 \pi \hbar v_\Dirac |\vecB| A / (P \Phi_0)$ yielding $\sim 4.4\,\textnormal{meV/T}$ (note that the expected subband spacing is approximately equal to 4.4~meV as well). As the trapping effect has an impact on several subbands on a much larger energy window (see Fig.~\nolink{\ref{fig:Deltaz}}b in Main Text), the Zeeman coupling can safely be neglected with respect to the orbital effect and related trapping effect, which is, in comparison, by far the dominant left-right symmetry breaking mechanism of the tri-junction.

\subsection{Planar Hall effect}
Due to the indirect coupling of an in-plane magnetic field with the surface state charge carriers via spin-polarized impurities, an in-plane magnetic field can break the symmetry in the transverse direction, inducing a Hall voltage in a 3D TI-based Hall bar, as reported in Ref.~\cite{Taskin17}. This effect is known as the planar Hall effect (PHE) and, interestingly, the PHE-induced Hall voltage profile has a similar $\pi$-periodicity as the one that we obtain for the steering ratio profile. This raises the question whether a PHE-induced transverse voltage profile in the input leg of the tri-junction can break the left-right symmetry of the transmission. When considering this explanation, the steering effect would not rely on a nanowire geometry for the legs and a magnetic field-induced trapping effect, but on a significant density of impurities on the 3D~TI surface. However, this explanation does not seem likely for the setup under consideration here, as the temperature dependence of the steering effect that we observe and the PHE are very different. In comparison to the steering effect, the PHE has a much weaker dependence on temperature and could be observed up to $200\,\textnormal{K}$. There is no pronounced cross-over temperature above which the PHE vanishes, but rather a steady linear decrease~\cite{Taskin17}.

\subsection{Weak antilocalization}
Weak antilocalization (WAL) affects the resistance of 3D~TI nanowires differently, depending on the relative orientation of the nanowire and the magnetic field, due to the difference in effective surface area for the surface state charge carriers in the plane perpendicular to the magnetic field~\cite{Koelzer20S}. The dependence of the resistance $R$ on this relative orientation is well described by  $R = R_\parallel + (R_\perp - R_\parallel) |\vece_\NW \cdot \vece_\vecB|$, with $\vece_\NW$ and $\vece_\vecB$ the unit vectors that represent the orientation of the nanowire and the magnetic field, respectively, and $R_\parallel$ ($R_\perp$) the resistance when the nanowire and the magnetic field have a parallel (perpendicular) alignment. Hence, WAL can break the left-right symmetry of the current through the tri-junction by affecting differently the resistances of the legs. However, due to the symmetric experimental setup of the T-junction device, with equal lengths between the junction and the contacts for all the legs, and the two output legs having the same orientation, WAL is expected to affect the total and individual currents minimally. This is in agreement with the isotropic profile of the total current as a function of the in-plane magnetic field components in Fig.~\nolink{\ref{fig:T-junction-current-map}}c in the Main Text. For the Y-junction, however, the experimental setup is less ideal. First, the Y-junction device did not allow for injection and extraction of the current at equal distances from the junction, due to broken contacts and, second, the left and right output legs are not aligned such that there is a slight asymmetry in how they are affected by WAL. Therefore, the T-junction current data displays a much clearer current map as compared to the Y-junction and allows for a clean extraction of the intrinsic steering ratio profile due to transmission across the tri-junction.

\begin{figure}
	\centering
	\includegraphics[width=.49\linewidth]{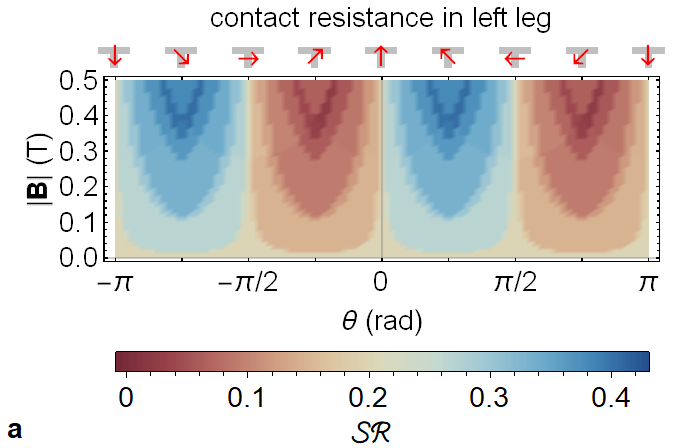}
	\includegraphics[width=.49\linewidth]{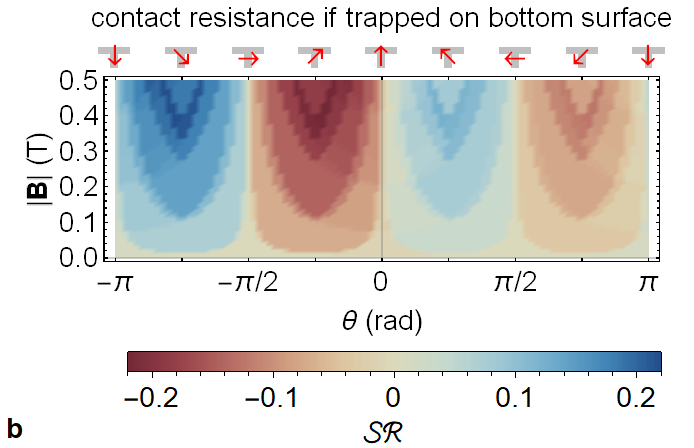}
	\caption{
	    (\textbf{a}),(\textbf{b}) The steering ratio is shown as a function of the in-plane magnetic field orientation angle $\theta$ and strength $|\vecB|$ up to $0.5\,\textnormal{T}$ for a T-junction with a contact resistance considered for states in the left output leg (in \textbf{a}), and for states that are trapped on the bottom surface (in \textbf{b}).
	}
	\label{fig:SR_T_Rc}
\end{figure}

\subsection{Asymmetry of the junction geometry and contact resistances}
The contact resistances of the different legs of the tri-junction can affect the current and steering ratio in two ways. First, an intrinsic difference in the contact resistances can be expected in general, such that an asymmetric current profile is already retrieved without an external magnetic field being applied. Second, the contact resistance can also be influenced by the trapping effect that is responsible for current steering across the junction. For example, there could be an increased contact resistance for a transport channel that is trapped on the bottom surface while the metal contact is applied to the top surface. These two aspects of contact resistances can be included in the transmission model in a straightforward manner by adjusting the transmission coefficients $T(l, k)$ to include a channel- and leg-dependent contact resistance $R^{(\cont)}(l, k)$:
\begin{equation}
	T(l, k) \rightarrow \frac{T(l, k)}{1 + G_0 R^{(\cont)}(l, k) T(l, k)},
\end{equation}
with $R^{(\cont)}(l, k)$ the total contact resistance for the channel under consideration, i.e., the sum of the contact resistances of the legs through which transmission is being considered.

In Supplementary Figs.~\ref{fig:SR_T_Rc}a and b, we modify the steering ratio profile as presented in Fig.~\nolink{\ref{fig:T-junction-B-constant}} of the Main Text by adding a contact resistance to the left output leg, and by applying a contact resistance to the transmission coefficient of a surface-state channel when it is trapped on the bottom surface, respectively. The former breaks the symmetry of the steering ratio profile around zero, introducing a net steering towards the right, while the latter breaks the $\pi$-periodicity of the steering ratio profile, as a positive (negative) angles $\theta$ traps the steering surface-state transport channels on the bottom (top) surface. The former type of symmetry breaking is generally observed in the experimental current data and is dealt with by subtracting the average steering ratio over all angles in the definition in Equation~\nolink{\eqref{eq:steering_ratio}} in the Main Text (omitted here to show the asymmetry explicitly), while the latter could not be observed.